\documentclass[
bibnotes,
amsmath,amssymb,
aps,
twocolumn,
prb,
]{revtex4}

\usepackage{graphicx}
\usepackage{dcolumn}
\usepackage{bm}
\usepackage{color}
\usepackage{lineno}

\def\la{~\mbox{\raisebox{-.6ex}{$\stackrel{<}{\sim}$}}~}
\def\ga{~\mbox{\raisebox{-.6ex}{$\stackrel{>}{\sim}$}}~}

\begin{document}

\title{Origin of magnetic interactions and their influence on the structural properties of Ni$_2$MnGa and 
        related compounds}


\author{Burak Himmetoglu, Vamshi M. Katukuri\footnote{present address:Leibniz Institute of Solid State and Materials Research, Dresden, Germany}, and Matteo Cococcioni}
 \affiliation{Department of Chemical Engineering and Materials Science, University of Minnesota, Minneapolis,
                Minnesota 55455}

\date{\today}
\begin{abstract}
In this work, we perform first principles DFT calculations to investigate the interplay between
magnetic and structural properties in Ni$_2$MnGa. We demonstrate that the relative stability of
austenite (cubic) and non-modulated martensite (tetragonal) phases depends critically on the magnetic
interactions between Mn atoms. While standard approximate DFT functionals stabilize the latter
phase, a more accurate treatment of electronic localization and magnetism, obtained with DFT+U,
suppresses the non-modulated tetragonal structure for the stoichiometric compound, in better agreement
with the experiments. 
We show that the Anderson impurity model, with Mn atoms
treated as magnetic impurities, can explain this
observation and that the fine balance between
super-exchange RKKY type interactions mediated by Ni $d$ and Ga $p$ orbitals
determines the equilibrium structure of the crystal.
The Anderson model is also demonstrated to capture the effect of the number 
of valence electrons per unit cell on the structural properties,
often used as an empirical parameter to tune the behavior of Ni$_2$MnGa based alloys.
Finally, we show that off-stoichiometric compositions with excess Mn promote transitions to a
non-modulated tetragonal structure, in agreement with experiments.
\end{abstract}

\maketitle

\section{\label{sec:Introduction}Introduction}
Magnetic Shape Memory alloys (MSMAs) 
are being widely explored 
for many technological applications. 
Typical representatives of this class of materials are
intermetallic Heusler alloys with X$_2$YZ composition 
(X and Y being transition metals).
The high temperature phase (Austenite) of these systems is usually cubic
and consists of four interpenetrating fcc lattices. At lower temperatures
they often assume a tetragonal structure (Martensite).
The transition between these phases is of martensitic type
and is usually characterized by high reversibility and tunability
~\cite{msm-rev,magn-1,magn-2,magn-3,magn-4,magn-5,magn-6,magn-7,magmech-1,magmech-2,
magmech-4,magmech-5}.
Being magnetic, these materials can show a strong coupling between
structural and magnetic transitions~\cite{magn-1,magn-2,magn-3,magn-4,
magn-5,magn-6,magn-7,magmech-4} which is very appealing for 
a wealth of different technological applications. 
In fact, Heusler alloys have been studied quite 
extensively for their magneto-caloric~\cite{magcal-1,magcal-2,magcal-3,magcal-4,magcal-5} and
magneto-mechanical~\cite{magmech-1,magmech-2,magmech-3,magmech-4,magmech-5} properties and more 
recently for their applications in energy conversion~\cite{econv}. 

Ni$_2$MnGa is a prototype representative of MSMAs. 
The stoichiometric alloy is ferromagnetic
below a Curie temperature of $T_C \simeq 365\, {\rm K}$~\cite{brown-modulated}.
The martensitic transition to a tetragonal phase, that takes place 
at $T_{M} \simeq 200\, {\rm K}$, preserves the ferromagnetic character, albeit
with a variation in the magnetic anisotropy. 
The martensitic phase is observed to have a modulated structure, 
where the parallel (110) planes are shifted from their equilibrium 
position along the [1${\bar 1}$0] direction 
with a period of five unit cells (5M) and a tetragonal distortion of $c/a \simeq 0.94$
~\cite{kokorin-modul,martynov-modulated,dai-eoa}. 
Under applied stress, an 
orthorhombic structure with a modulation of seven unit cells (7M) has also been observed
~\cite{brown-modulated,kaufmann-modulated,kokorin-modul,wedel-modulated}. 
Instead, a non-modulated, tetragonal martensite with $c/a \simeq 1.2$ can be obtained
for off-stoichiometric compositions at lower temperatures~\cite{kokorin-modul,sozinov-aniso} 
($T \la 193\, K$). Several groups have also claimed 
the existence of a pre-martensitic phase below a transition temperature of $T_{\rm PM} \simeq 260\, {\rm K}$,
where the structure is modulated with a period of three unit cells (3M)~\cite{brown-modulated}.
Co-existence of several modulated and non-modulated phases in a twinned structure
has also been observed experimentally~\cite{kaufmann-modulated}.

Ni$_2$MnGa and related Heusler alloys have been studied computationally in several works that  
investigated their electronic and structural properties. 
These studies, mostly based on the local density (LDA) or the 
generalized gradient approximation (GGA) for the 
exchange-correlation functional, have been successful in describing the electronic,
structural~\cite{enkovaara-coa,entel-review,rabe-coa,ayuela-coa,ayuela-coa2,kart-coa,enkovaara-coa2,hu-x,Ni2MnSn},
vibrational properties~\cite{rabe-vib,rabe-phon,el-ph}, and computing the phase diagram~\cite{entel-review,phase} and
the Curie temperature~\cite{heusler-rkky2,heusler-rkky4,Tc} for these types of compounds.
A particularly well studied aspect
is the relative stability of the cubic austenite and a non-modulated
martensite phases with $c/a \neq 1$, 
which has been investigated by first-principles calculations using various functionals
~\cite{enkovaara-coa,entel-review,rabe-coa,ayuela-coa,ayuela-coa2,kart-coa,enkovaara-coa2}. 
With few exceptions, these studies have yielded 
qualitatively similar results that, while in disagreement on the amount of
the tetragonal distortion, generally predict a stable non modulated
martensite phase with $c/a > 1$.
Several mechanisms have been proposed as responsible for the structural distortions of this material, 
such as the Jahn-Teller effect~\cite{brown-jahnteller,ayuela-coa2,fujii-jahnteller} and Fermi surface
nesting, which was found to be related to phonon mode softening~\cite{opeil-pre,rabe-phon,zheludev-fs}. 
It was also found that the number of valence electrons per atom ($e/a$) can control the relative 
stability of different phases, the martensitic transition temperature and the softening of the phonon modes.
~\cite{rabe-eoa,hu-eoa,dai-eoa}. 
However, the microscopic mechanisms responsible for these effects have not yet been completely understood.
As a consequence, engineering these materials for specific applications relies on the empirical optimization 
of alloy composition and is mostly based on a trial and error approach. 
Experiments have also found that the samples in which a non-modulated martensitic phase is stable are
off-stoichiometric ones, with an excess Mn content in their compositions
~\cite{sozinov-aniso,lanska-coa,magmech-5,dai-eoa}. 
In fact, a related compound Ni$_2$MnSn has been shown
to have a low temperature tetragonal structure with $c/a > 1$ only with a Mn content 
larger than the stoichiometric composition, both experimentally and by first-principles calculations
~\cite{Ni2MnSn}.

In this work, we perform a detailed study of the relative stability of the cubic and 
non-modulated martensitic structures and try to clarify some of the conflicting results in the literature.
After computing the effective exchange interaction parameters between Mn atoms with constrained density functional
theory (DFT), we determine an approximate magnetic energy using the Heisenberg model. 
We show that the difference in energy between the austenite and martensite phases can be accounted for 
by the variation of Heisenberg magnetic coupling.
By using the DFT+U functional~\cite{Ucalc, anisimov-1991, anisimov-1993, mazin-1997, solovyev-1998}, 
we also demonstrate that electronic correlations 
play an important role in determining the effective exchange interactions.
In particular, DFT+U suppresses 
the stability of the non-modulated martensitic structure for stoichiometric compositions. 
These results are interpreted using the Anderson 
impurity model~\cite{anderson-imp}, where Mn atoms are treated as a periodic array of magnetic impurities
embedded in a conduction electron gas composed of electrons on Ni $d$ and Ga $p$ orbitals, 
which mediate super-exchange
interactions between Mn atoms. 
Using this idea, we are able to predict the strength of the super-exchange interactions,
and to show that the balance
between the super-exchange through Ga $p$ and Ni $d$ orbitals
determines the stability of the structure.
We also study the effect of the number of electrons on the effective exchange 
interactions, by injecting electrons into the unit cell. We show that the number of injected electrons
affects the Fermi surface, which in turn modifies 
the strength of super-exchange interactions and
the stability of the martensitic phase. Finally, we show that the martensitic phase can be
stabilized when the Mn content in the material exceeds the stoichiometric composition value,
which is in agreement with the experimental observations. 

This paper is organized as follows: In section~\ref{sec:Method}, we summarize the computational methods we have employed.
In section~\ref{sec:GGA}, we discuss calculations performed in the austenite and martensite phases, and discuss their
relative stability. In section~\ref{sec:exchange}, we present the calculation of effective exchange parameters
between Mn atoms. In section~\ref{sec:mag}, we discuss the underlying magnetic coupling mechanisms between Mn atoms.
In section~\ref{sec:extension}, we discuss the effects of the 
number of electrons in the unit cell and the extra Mn impurities
on the structural properties. Finally in section~\ref{sec:conclusion}, we propose some conclusive remarks.

\section{\label{sec:Method}Computational Method}
Calculations in this paper were performed using the plane-waves pseudopotential implementation of DFT contained 
in the ``pwscf'' code of the {\it Quantum ESPRESSO} package~\cite{espresso}.
The generalized gradient approximation (GGA) functional with the Perdew-Burke-Ernzherof (PBE) parametrization~\cite{pbe}
was employed for the exchange-correlation energy.
The Ni, Mn and Ga atoms were all represented by ultrasoft pseudopotentials~\cite{vanderbilt}.
The electronic wavefunctions and charge density were expanded up to kinetic energy cut-offs of
$45\, {\rm Ry}$ and $480\, {\rm Ry}$ respectively.
The Brillouin zone integrations were performed using a $12\times 12\times 12$ Monkhorst and
Pack special point grids~\cite{BZ} and a Methfessel and Paxton smearing of the Fermi-Dirac distribution
function~\cite{mp}, with a smearing width of $0.01\, {\rm Ry}$. 

In order to improve the description of electronic correlation and localization,
the Hubbard-model-based DFT+U corrective functional was employed.
The on-site Coulomb repulsion parameters were computed using 
the linear response method introduced in Ref.~\onlinecite{Ucalc}. The $s$ states of Ni, Mn, and Ga
were treated as a charge ``reservoir'' as described in Ref.~\onlinecite{hwc}. The linear response calculation of $U$ 
provides the intersite interaction parameters $(V)$~\cite{HubV} as well; however, we found that they are significantly 
smaller than the on-site interaction parameters $(U)$, therefore we neglect them in the calculations presented
in this paper. 
The crystal structures presented in this paper were generated using the XCrysDen software~\cite{xcrysden}.

\section{\label{sec:GGA}GGA and GGA+U Calculations}
\subsection{\label{sub:DOS}Cubic phase}
\begin{figure}[!ht]
\begin{minipage}[b]{0.45\linewidth}
\centering
\includegraphics[width=1.1\textwidth]{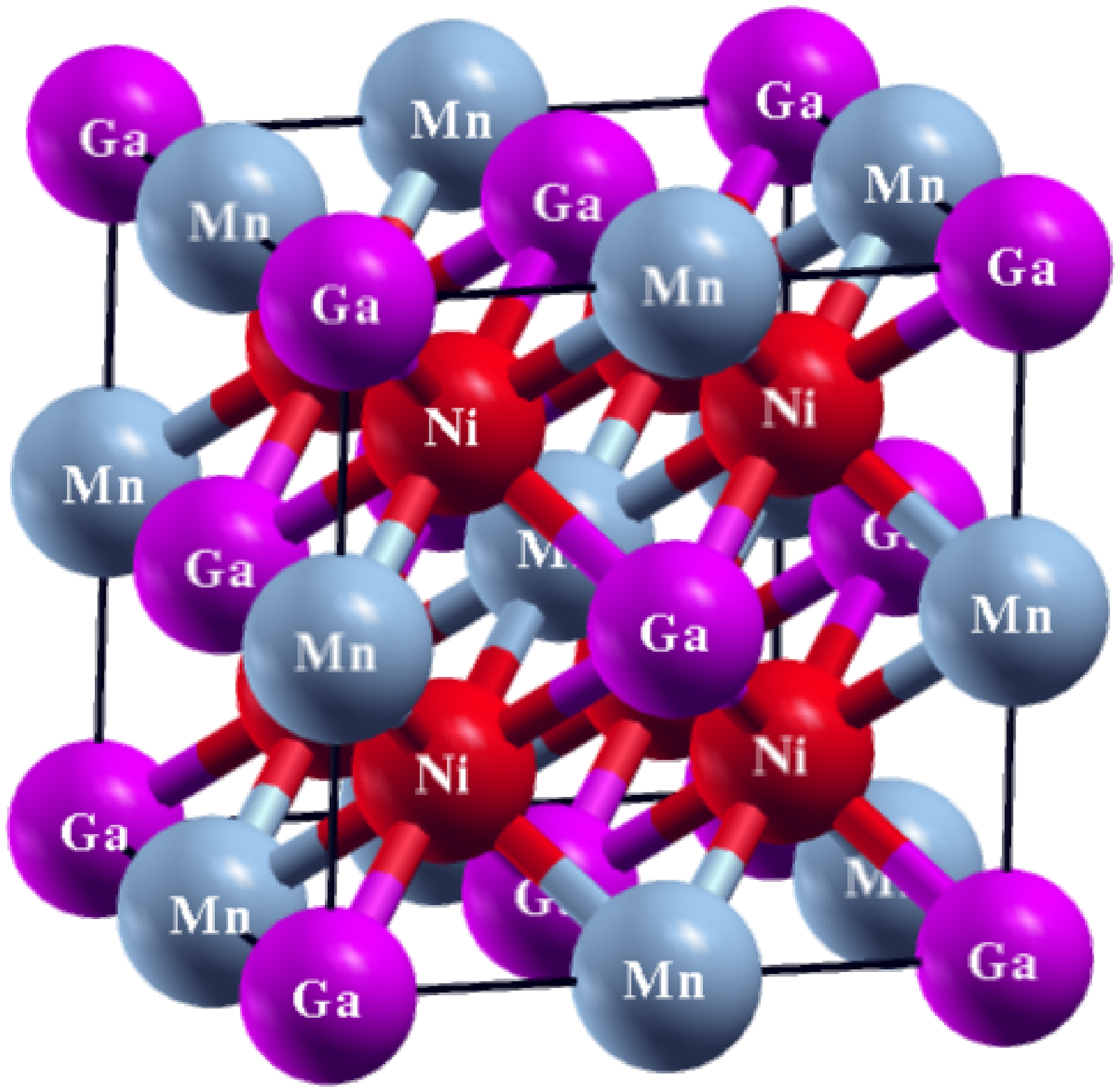}
\end{minipage}
\hspace{0.4cm}
\begin{minipage}[b]{0.45\linewidth}
\centering
\includegraphics[width=1.1\textwidth]{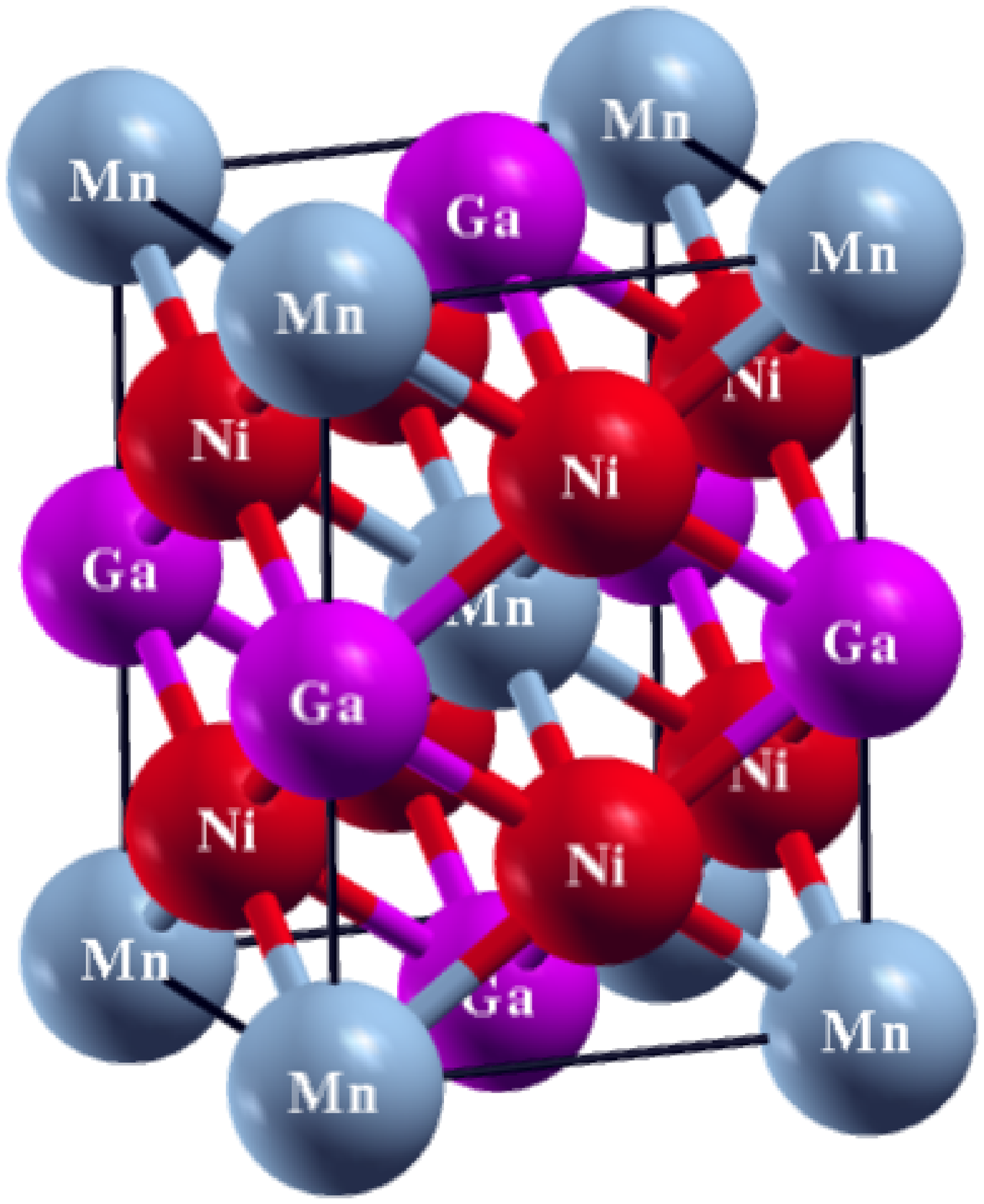}
\end{minipage}
\caption{\label{fig:cells} (Color online) The $L2_1$ structure is shown on the left, and the tetragonal unit cell
is shown on the right.  Both unit cells correspond to the cubic phase.}
\end{figure}

The high-temperature $(T>T_M)$ cubic phase of Ni$_2$MnGa has $L2_1$ 
structure where the Ni, Mn and Ga atoms
form four inter-penetrating fcc lattices (two of which are occupied by Ni), as shown in Fig.~\ref{fig:cells}. 
Also shown in Fig.~\ref{fig:cells},
is the tetragonal unit cell, whose basal unit vectors are parallel to $[110]$ directions of the $L2_1$
structure.
To describe the tetragonal deformation of the unit cell, the ratio $c/a$ is generally used as
measured from the cubic cell ($c/a=1$). The equilibrium lattice parameter and magnetization determined
with GGA calculations reported in Table~\ref{tab:gga} are in good agreement with experiments~\cite{heusler}.
\begin{table}[b]
\caption{\label{tab:gga}
Calculated lattice parameter and magnetizations of each atom in the cubic phase.}
\begin{ruledtabular}
\begin{tabular}{cccccc}
 & $a_0\,$ (\AA)  & $\mu_{\rm Mn}\, (\mu_B)$ & $\mu_{\rm Ni}\, (\mu_B)$ & $\mu_{\rm Ga}\, (\mu_B)$ &  
$\mu_{\rm tot}\, (\mu_B/{\rm cell})$   \\
\hline
GGA & $5.83$ & $3.67$ & $0.34$ & $-0.13$ & $4.22$ \\
GGA+U & $5.83$\footnotemark[1] & $4.52$ & $0.16$ & $-0.13$ & $4.80$ \\
\end{tabular}
\end{ruledtabular}
\footnotetext[1]{Kept at the GGA calculated value.}
\end{table}
%
%
\begin{figure}[!ht]
\includegraphics[width=0.45\textwidth]{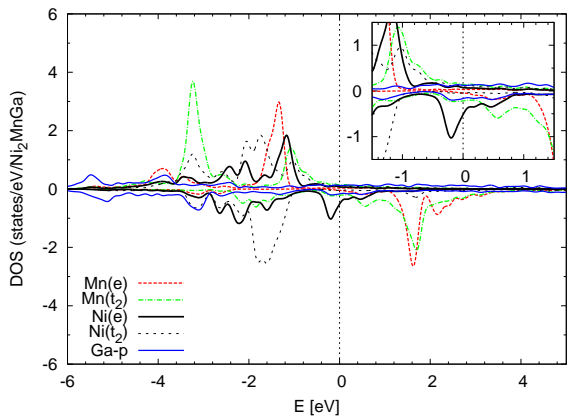}
\caption{\label{fig:pdos.gga} (Color online) The projected density of states calculated using the GGA functional.
The Fermi level is set to the zero of energy.
The inset shows the density of states in the close proximity of the Fermi level.}
\end{figure}
Our calculations show that the Mn $d$ states 
split into two groups of three and two degenerate states, that according 
to tetrahedral group notation, are indicated by $t_2$ and $e$ respectively, 
whose density of states, obtained with the GGA functional are shown
in Fig.~\ref{fig:pdos.gga}.
This shows that the Mn $d$ states are subject to a cubic crystal field.
More precisely, with respect to the 
Cartesian coordinates on the tetragonal unit cell (when $c/a=1$) in Fig.~\ref{fig:cells},
$d_{xy}$ and $d_{z^2}$ orbitals of Mn are the $e$ states and 
$d_{zx},\, d_{zy}, \, d_{x^2-y^2}$ orbitals are the $t_2$ states. 
Similarly, the Ni $d$ orbitals are also split into $e$ and $t_2$ states,
with $d_{xy}$ and $d_{z^2}$ orbitals being the $e$ and $d_{zx}, \, d_{zy}, \, d_{x^2-y^2}$ orbitals being 
the $t_2$ states. 

Another feature clearly visible in the density of states is that the magnetization density is mainly localized
on the Mn $d$ orbitals. The peaks corresponding to majority and minority  states are clearly
separated, with minority spin states nearly empty.
The localized nature of magnetization on the Mn $d$ orbitals suggest that on-site Coulomb repulsion
could play a dominant role. It can also be expected that GGA functionals would not be able to account 
for this interaction accurately, since they tend to delocalize the electronic density. To improve the
description of localized electrons responsible for magnetism, we adopted the DFT+U approach
~\cite{Ucalc, anisimov-1991, anisimov-1993, mazin-1997, solovyev-1998}.

\begin{figure}[!ht]
\includegraphics[width=0.45\textwidth]{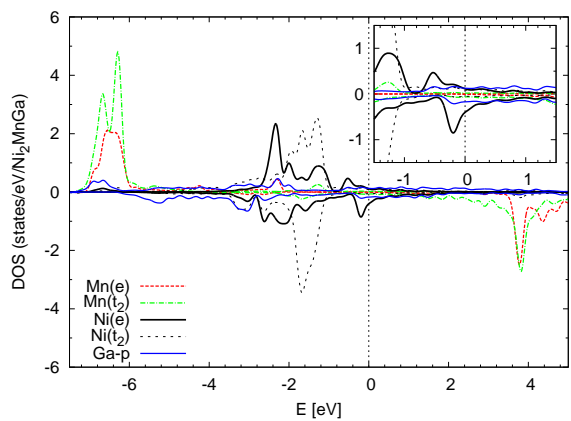}
\caption{\label{fig:pdos.u} (Color online) The projected density of states calculated using the GGA+U functional.
The Fermi level is set to the zero of the energy.
The inset shows the density of states in the close proximity of the Fermi level.}
\end{figure}

The GGA+U calculations were performed by using $U = 5.97\, {\rm eV}$ on the $d$ states of Mn. The value of
the $U$ correction is calculated from the linear response approach of Ref~\onlinecite{Ucalc}, in a $16$ atom super cell 
(four times larger than the unit cell). 
We have explicitly verified that the calculated Hubbard parameters are unchanged when larger
supercells are used, as explained in Ref.~\onlinecite{Ucalc}. Although $U = 5.97\, {\rm eV}$ might seem
unexpectedly large for a metallic system, it is used for describing localized electrons on Mn $d$ states, 
which have vanishing contribution to the density of states 
at the Fermi level.
On the contrary, the metallic character is mainly due to Ni $d$ ($e$) and Ga $p$ 
states that dominate the density of states at the Fermi level.
%
%
The $U$ on Ni $d$ and Ga $p$ states, as well as the interaction terms between electrons 
in different manifolds (e.g., $d$ and $s$ states of Mn) 
or on different sites (inter-site interactions $V$~\cite{HubV}) are neglected in this work as they are found to be 
largely irrelevant for the conclusions of this work. Also, the calculation of the electronic parameters 
was not performed self-consistently as in Ref.~\onlinecite{self-cons-u}, nor they were 
recomputed for different distortions of 
the unit cell~\cite{str-cons-u}, as their variation with the crystal structure was found to be unimportant.
The density of states obtained from the GGA+U
functional is shown in Fig.~\ref{fig:pdos.u}. The most significant difference from the density of 
states calculated with the GGA functional in Fig.~\ref{fig:pdos.gga}, is the increase in the splitting 
between the majority and minority spin peaks of the Mn $d$ states 
(consistent with other DFT+U calculations),
and their complete disappearance from the Fermi level.
At this point, we would also like to comment on the photo-emission type experiments performed
in the literature, which could be compared to the calculated density of states. These studies have
become available relatively recently on Ni$_2$MnGa(Sn)~\cite{opeil-pre,xray,Ni2MnSn} and the data
presented were able to provide information mainly on states close to the Fermi level. These
states are mainly of Ni $e$ type, and as Refs.~\onlinecite{opeil-pre,Ni2MnSn} also suggest, they are
responsible for the nesting of the Fermi surface. 
Our GGA+U calculations modify almost entirely Mn $d$ states,
pushing them to lower energies, where experiments are not conclusive. Instead, the $U$ correction has
little effect on the states at the Fermi level, and therefore our results are still consistent with existing data.

The more complete filling of electrons on the majority spin states of Mn leads to a 
larger magnetization, $4.80\, {\mu}_B$ per unit cell. 
The magnetization of each atom, computed with the GGA+U functional, is shown in Table~\ref{tab:gga}.
In our calculations, we fixed the cubic lattice parameter
to the value obtained from the GGA functional for simplicity. 
As can be seen from Table~\ref{tab:gga}, 
while the magnetization of Ga remains at the same value, the magnetization of Mn increases  
and the magnetization of Ni decreases compared to GGA. 
The GGA+U results are in overall agreement with the
neutron scattering data~\cite{brown-jahnteller}, even if the Mn magnetization is slightly overestimated.
In this regards, it should be noted that the extraction of the 
magnetic moment of each atom from the data requires a priori
assumptions, which affect the overall outcome of the analysis, and leads to large uncertainties.
We can obtain a better understanding of the
effects of the Hubbard correction, by considering the trace of the occupation matrices 
$n_{m\, m'}^{I\, \sigma}$ defined in Ref.~\onlinecite{Ucalc}, which is a measure of how many electrons 
localize on a site $I$ with spin $\sigma$ (while $n^I = \sum_{m,\sigma} n^{I\, \sigma}_{m\, m}$ is  
the total number of electrons on site $I$).
In the GGA ground state, we found that $n_{Ni-d} \simeq 8.83$, 
$n_{Mn-d} \simeq 5.47$ and $n_{Ga-p} \simeq 2.43$ while in GGA+U, $n_{Ni-d} \simeq 8.90$, 
$n_{Mn-d} \simeq 5.21$ and $n_{Ga-p} \simeq 2.46$.
In the GGA ground state, there are $n_{Mn-d}^{\downarrow} \simeq 0.9$ minority spin
electrons shared between the $t_2$ states of Mn, compared to $n_{Mn-d}^{\uparrow} \simeq 4.57$ of 
majority spin. In GGA+U, which favors integer occupation of orbitals, 
the minority spin $d$ states are almost empty with $n_{Mn-d}^{\downarrow} \simeq 0.34$, while 
$n_{Mn-d}^{\uparrow} \simeq 4.86$ for majority spin, closer to being fully occupied.
While some of the minority spin electrons in Mn $t_2$ states are 
transferred to the majority spin states in GGA+U, the rest mostly contribute to the conduction electrons
of Ni $d$ type (since increase in Ga $p$ occupation in GGA+U is small).
This electronic re-organization in the GGA+U ground state has important consequences
in the structural properties of the material, as shown in the next section. 
\subsection{Tetragonal distortions}
\begin{figure}[!ht]
\includegraphics[width=0.45\textwidth]{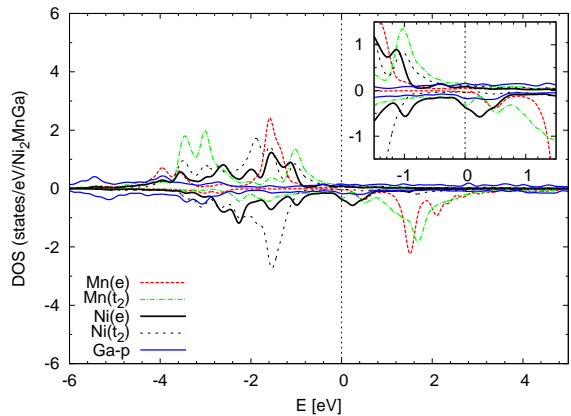}
\caption{\label{fig:pdos.1.2.gga} (Color online) The projected density of states calculated using the GGA functional
at $c/a = 1.2$. The Fermi level is set to the zero of energy.
The inset shows the density of states in the close proximity of the Fermi level.}
\end{figure}

\begin{figure}[!ht]
\includegraphics[width=0.45\textwidth]{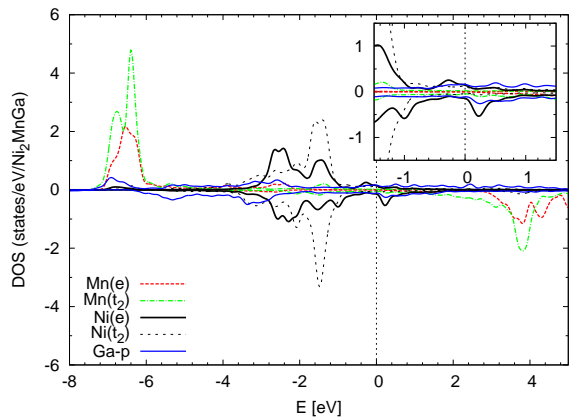}
\caption{\label{fig:pdos.1.2.u} (Color online) The projected density of states calculated using the GGA+U functional
at $c/a=1.2$. The Fermi level is set to the zero of energy.
The inset shows the density of states in the close proximity of the Fermi level.}
\end{figure}

We first discuss the electronic structure of the low-temperature phase of Ni$_2$MnGa, 
with an experimental tetragonal distortion of $c/a \simeq 1.2$~\cite{kokorin-modul,sozinov-aniso}. 
This phase has often been studied in the literature for the stoichiometric
compound, as many calculations obtain it as a ground state.  
In order to investigate the electronic re-organizations when the unit cell is tetragonally distorted at 
constant volume, we have computed the density of states using both the GGA (shown in 
Fig.~\ref{fig:pdos.1.2.gga}) and the GGA+U (shown in 
Fig.~\ref{fig:pdos.1.2.u}) at $c/a = 1.2$.  
The value of $U$ was set to the calculated value from the cubic phase. 
We have verified the validity of this approximation by recomputing $U$ for various tetragonal distortions
(at constant volume), and we found that $U$ is almost constant. 

\begin{figure}[!ht]
\includegraphics[width=0.45\textwidth]{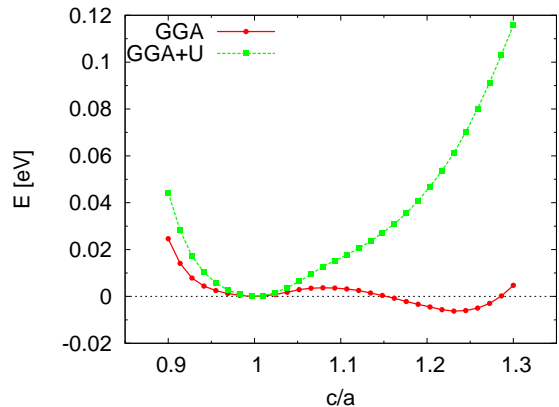}
\caption{\label{fig:evscoa} (Color online) Energy of the ground state per unit cell as a function of $c/a$ at constant volume
in GGA and GGA+U. The zero of energy is set to be at $c/a=1$ in each case.}
\end{figure}

In the tetragonal cell, Mn and Ni $t_2$ and $e$ orbitals split (the $t_2$ orbitals split into a 
doublet $d_{zx}$, $d_{zy}$ and a singlet $d_{x^2-y^2}$, while the $e$ splits into two singlets)
as can be seen from Figs.~\ref{fig:pdos.1.2.gga} and~\ref{fig:pdos.1.2.u}. Some previous
works in the literature relate the existence of the tetragonal phase at $c/a \simeq 1.2$ to a Jahn-Teller
type distortion~\cite{brown-jahnteller,ayuela-coa2,fujii-jahnteller}. 
From the inset in Fig.~\ref{fig:pdos.1.2.gga}, we observe that the Ni minority spin $e$ peak, 
which represents the dominant contribution at the Fermi level splits, forming a shoulder towards the lower
energy direction. This shoulder is due to the lower energy $d_{z^2}$ orbitals, while $d_{xy}$ orbitals are
pushed to higher energy. 
However, peaks corresponding to both types of orbitals move above the Fermi 
level while maintaining a tail below it. This feature is much stronger in GGA+U, where both minority spin Ni $e$ 
orbitals move entirely above the Fermi level, as shown in the inset of Fig.~\ref{fig:pdos.1.2.u}.
Jahn-Teller type distortions would tend to place the Fermi level between the split orbitals, resulting in 
a fully occupied lower energy and an empty higher energy orbital. Instead, we observe that the split orbitals
both move above the Fermi level. Therefore, we argue that another mechanism
is responsible for determining the energy landscape of the material and in particular, the stability of the 
tetragonal phase. In order to investigate this point further, we have performed total energy calculations 
in GGA and GGA+U as a function of the tetragonal distortion ($c/a$) at constant volume, 
which are shown in Fig.~\ref{fig:evscoa}.
As can be seen, the GGA functional yields a local minimum at $c/a=1$ and a global minimum at $c/a \simeq 1.23$. 
The energy difference between the two minima is very small and is around $4\, {\rm meV}$. Numerous studies
in the literature have previously found similar results to our GGA calculation
~\cite{enkovaara-coa,entel-review,rabe-coa,ayuela-coa,kart-coa}. However, these studies have
identified energy minima at slightly different values of $c/a$, and proposed different explanations about the 
relative stability of different phases. 
For example, Ref.~\onlinecite{rabe-coa} has identified a local 
energy minimum for a non-modulated orthorhombic structure at $c/a \simeq 1.11$ and $b/a \simeq 1.04$, 
besides the local minimum at $c/a=1$ and a global minimum at $c/a \simeq 1.2$. 
Ref.~\onlinecite{enkovaara-coa}
has found a local energy minimum at $c/a \simeq 0.94$ and a global energy minimum at $c/a \simeq 1.3$ for 
the non-modulated structure, while Ref.~\onlinecite{enkovaara-coa2} found a local energy minimum at $c/a \simeq 0.94$
for the off-stoichiometric non-modulated compound Ni$_2$Mn$_{1.25}$Ga$_{0.75}$.
Instead, 
Refs.~\onlinecite{kart-coa,entel-review,ayuela-coa} have found that an energy minimum for $c/a<1$ is only 
possible for modulated structures. 
Experimentally, a distortion with $c/a<1$ seems to correlate with the modulation of the structure
~\cite{wedel-modulated,brown-modulated,martynov-modulated,kaufmann-modulated}. 
A tetragonal non-modulated phase seems to be stable only for significant deviations from the 
stoichiometric compositions~\cite{sozinov-aniso,lanska-coa,magmech-5,dai-eoa}.
Therefore, experiments seem to not support the existence of a non-modulated 
tetragonal phase at the stoichiometric
composition, which is in contrast with the result of the GGA calculation. 

\begin{figure}[!ht]
\includegraphics[width=0.45\textwidth]{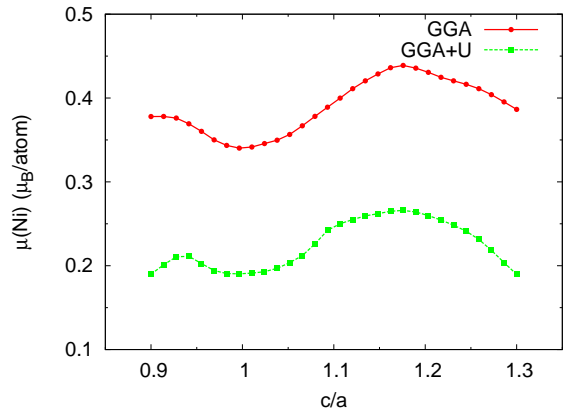}
\caption{\label{fig:magNi} (Color online) Magnetization of Ni atoms as a function of $c/a$ computed
using GGA and GGA+U.}
\end{figure}

\begin{figure}[!ht]
\includegraphics[width=0.45\textwidth]{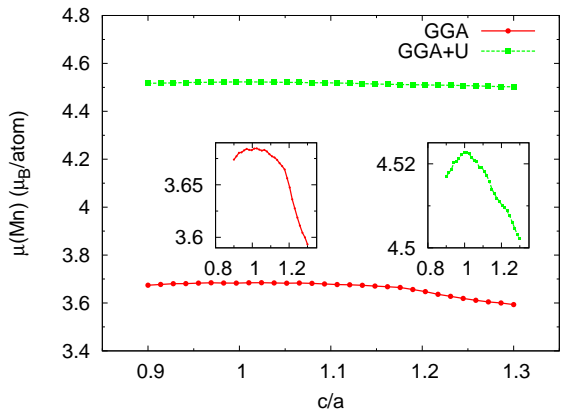}
\caption{\label{fig:magMn} (Color online) Magnetization of Mn atoms as a function of $c/a$ computed
using GGA and GGA+U. The insets show a close up view of the magnetizations separately.} 
\end{figure}

\begin{figure}[!ht]
\includegraphics[width=0.45\textwidth]{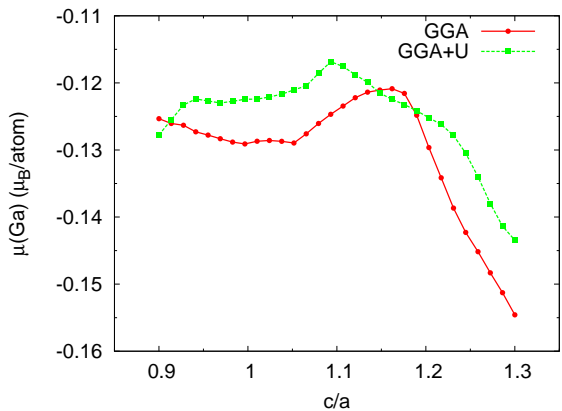}
\caption{\label{fig:magGa} (Color online) Magnetization of Ga atoms as a function of $c/a$ computed
using GGA and GGA+U.}
\end{figure}

In GGA+U, the global energy minimum at $c/a \simeq 1.23$ is suppressed as shown in Fig.~\ref{fig:evscoa}.
This finding indicates that obtaining a more accurate description of 
electronic localization on $d$ states is important to predict the relative stability of 
different phases. The suppression of the stable tetragonal phase shows that GGA+U is in better 
agreement with experiments, compared to GGA. 
In order to understand the relative energies of different phases, and their relation to 
magnetism in GGA and GGA+U, we have computed the magnetizations of each atom as a function of $c/a$, and the results 
are reported in Figs.~\ref{fig:magNi},~\ref{fig:magMn},~\ref{fig:magGa}.
An interesting feature that is apparent from Fig.~\ref{fig:magNi} is that the energy minima in the GGA 
functional corresponds approximately to extrema of magnetization of the Ni atom (i.e. the minimum magnetization of 
Ni is obtained at $c/a = 1$, while the maximum magnetization is obtained at $c/a \simeq 1.2$).
On the other hand, the magnitude of the magnetization of Ga atoms has the lowest value around 
$ 1.1 \la c/a \la 1.2$
but increases monotonically for $c/a > 1.2$. The Mn atoms have the highest magnetization at $c/a = 1$,  
and it decreases monotonically for $c/a > 1.2$, with a rather featureless profile.
Therefore, the magnetizations of Ni and Ga atoms
seem to correlate with opposite minima in the total energy as a function of $c/a$. 
The oscillatory variation of magnetization of Ni atoms suggests a correlation with electronic screening, 
since the dominant contribution to the Fermi surface comes from the minority spin electrons of Ni atoms. 
It is well known that, when an impurity is embedded in a gas of conduction electrons, 
its potential is screened and leads to an
oscillatory variation of the charge density of free electrons, whose period depends on the Fermi momentum
(Friedel oscillations). 
If Mn atoms are treated as a periodic array of ``impurities'' and the Ni $d$ states as the conduction electron 
sea, a similar oscillatory behavior can be expected. Since the minority spin $e$ electrons of Ni 
provide the dominant 
contribution to the density of states at the Fermi level, they contribute to the screening of the impurity potential
most significantly.
According to this picture, 
the oscillations in the magnetization of Ni $d$ states can be thought of as due to spin-polarized Friedel 
oscillations.
Notice that the oscillations are with respect to $c/a$ and 
not the distance between the Ni atoms. As $c/a$ increases, the distance between Ni atoms decreases in the 
basal plane, while the vertical distance between them increases, which introduce a more complicated 
dependence of the oscillations on $c/a$ than in the free electron case (through the anisotropies  
of the Fermi surface). 
We therefore treat these oscillations at a qualitative level. 
\section{\label{sec:exchange}Exchange Interaction between ${\rm Mn}$ atoms}
\begin{figure}[!ht]
\includegraphics[width=0.45\textwidth]{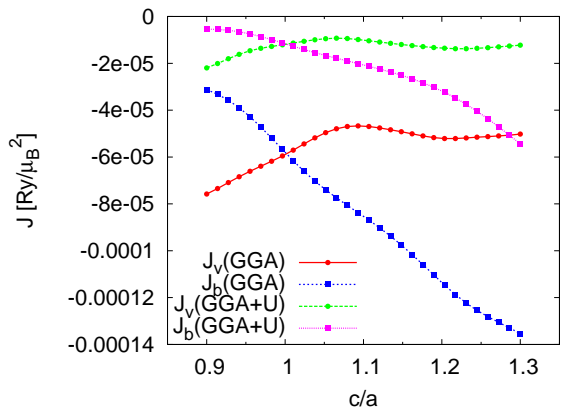}
\caption{\label{fig:jall} (Color online) Exchange parameters $J_v$ and $J_b$ computed using GGA and GGA+U 
functionals.}
\end{figure}

The ferromagnetic ordering is a consequence of the magnetic interactions in the system,
mainly between Mn atoms.
Therefore, we can approximate the magnetic interaction energy through the Heisenberg Hamiltonian, 
that depends only on the magnetization of Mn atoms:
\begin{equation}
{\cal H}_{\rm mag} = \sum_{\langle i, \, j \rangle} J_{i, j}\, {\bf S}_i \cdot {\bf S}_j . \label{heis}
\end{equation}
In Eq.(\ref{heis}) $J_{i, j}$ are the exchange parameters, ${\bf S}_i$'s are magnetizations of Mn atoms and
$\langle i,\, j \rangle$ denotes that the sum runs over nearest neighbor Mn atoms.
This approximation is justified by our expectation that
the strength of the exchange parameter decays rapidly with distance.
In order to compute the exchange parameters $J_{i,j}$, we use constrained DFT calculations in a 16 atom
supercell, that contains four Mn atoms.
In this calculation, the magnetic moment of one of the Mn atoms is flipped to an opposite direction with 
respect to the rest of the Mn atoms (while its magnitude turns out to remain unchanged). 
The energy difference with the original ferromagnetic configuration is then mapped on the Heisenberg 
model of Eq.(\ref{heis}) to obtain $J_{i,j}$.
\begin{figure}[!ht]
\includegraphics[width=0.4\textwidth]{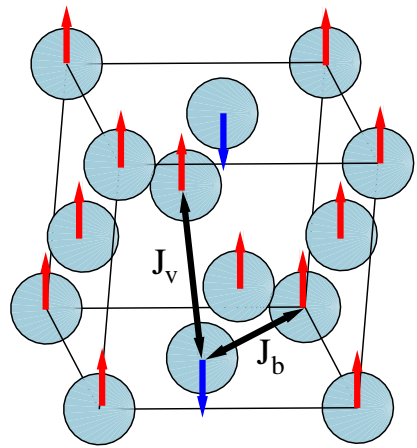}
\caption{\label{fig:mncell} (Color online) Mn atoms in the supercell used to compute the exchange parameters.}
\end{figure}
This calculation is performed for a range of $c/a$ values. For the tetragonal cells,
there are two relevant exchange parameters; one
in the $x-y$ plane ($J_b$) between the Mn atoms connected in the $[110]$ direction
and one in the $y-z$ plane ($J_v$) between the Mn atoms connected in the $[011]$ direction as shown in
Fig~\ref{fig:mncell}. Naturally, the two exchange parameters $J_v$ and $J_b$ are equal in the cubic limit 
($J_v = J_b = J_{\rm cubic}$ at $c/a=1$). The magnetic energy per supercell is given by 
\begin{equation}
{\cal H}_{\rm mag} = \left( 4 J_b + 8 J_v \right)\, \left( S_{+}^2 + S_{+}\, S_{-} \right) \label{Hmag}
\end{equation}
where $S_{+}$ is the magnetization of the spin up Mn atoms and $S_{-}$ is the magnetization of the flipped 
spin Mn atom. The calculation is performed by first fixing the basal plane dimension $a$ 
(thus, $J_b=J_{\rm cubic}$) and 
varying $c$ to determine $J_v$ as a function of $c$ using Eq.(\ref{Hmag}).
Then, $c$ is fixed (thus, $J_v=J_{\rm cubic}$) 
and $a$ is varied to determine $J_b$ as a function of $a$. The values of 
$c$ and $a$ for each constrained DFT calculation are chosen such that $a^2\, c = {\rm constant}$ and therefore
both $J_v$ and $J_b$ can be expressed as a function of $c/a$ (corresponding to constant volume deformations)
only.
The resulting values of $J_v$ and $J_b$ as a function of $c/a$ are shown in Fig.~\ref{fig:jall}
for both the GGA and GGA+U functionals.
As can be seen from Fig.~\ref{fig:jall}, the magnitude of the exchange parameter $J_b$ increases with $c/a$,
while that of $J_v$ decreases until $c/a \sim 1.1$ after which it plateaus.  
This is expected, since for larger $c/a$ the distance between Mn atoms in the $x-y$
plane decreases resulting in a stronger magnetic interaction in the plane ($J_b$), while the distance between Mn atoms
in the $y-z$ and $x-z$ plane increases, resulting in a weaker interaction ($J_v$). The plateau in $J_v$ is 
probably related to the vertical alignment of Mn atoms that partially compensates for the
increased distance between $(001)$ planes. The dependence shown in Fig.~\ref{fig:jall} also justifies the
energy minimum at $c/a > 1$ for GGA. 
The stronger increase in magnitude of $J_b$ compared to $J_v$ indicates 
that the magnitude of the magnetic energy in Eq.(\ref{Hmag}) increases with $c/a$ 
(since the increase in absolute magnetic energy due to $J_b$ dominates the decrease due to $J_v$).
Instead, for $c/a < 1$, the increase in $J_b$ is
almost symmetrically cancelled by a decrease in $J_v$ so no energy minimum is observed. 

Another important result we obtain is that the magnitude of 
both interaction parameters are significantly smaller in GGA+U than in GGA. 
In the simple Hubbard model, the exchange interaction between electrons localized on neighboring atomic sites 
can be determined from 2nd order perturbation theory as 
$J \sim t^2/U$, where $t$ is the hopping amplitude between sites~\cite{anderson-tJ,spalek-tJ}.
In light of this fact, smaller exchange parameters between Mn atoms
in GGA+U are expected, since the effective on-site Coulomb repulsion is larger.
Notice that the $J$ in the simple Hubbard model is positive indicating an anti-ferromagnetic interaction, while
in our constrained DFT calculations we find negative $J$'s, resulting in a ferromagnetic ground state. 
This discrepancy is due to a different type of mechanism that leads to magnetism in our 
system compared to the simple Hubbard model. We will discuss this point in detail in the next section.
Notice also that both exchange parameters $J_v$ and $J_b$ have an oscillatory modulation, as was the case
for the magnetization of Ni $d$ states shown in Fig.~\ref{fig:magNi}. This
modulation seems to suggest that Ni $d$ states contribute to the mediation of magnetic interactions between
Mn atoms. Indeed, the oscillation of the exchange parameters between localized 
magnetic impurities in a free electron gas is a well known effect~\cite{rkky}. 
We postpone the study of possible mechanisms until the
next section, where we will provide a detailed discussion about the nature of the magnetic interaction between 
the Mn atoms. In the rest of this section, we study the effect of the magnetic energy on the structural properties
of the material. 
\begin{figure}[!ht]
\includegraphics[width=0.45\textwidth]{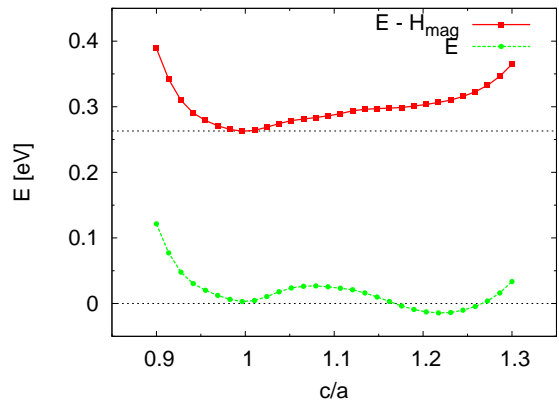}
\caption{\label{fig:emag.gga} (Color online) Comparison of the total ground state energy and the magnetic energy
per unit cell in GGA.}
\end{figure}
\begin{figure}[!ht]
\includegraphics[width=0.45\textwidth]{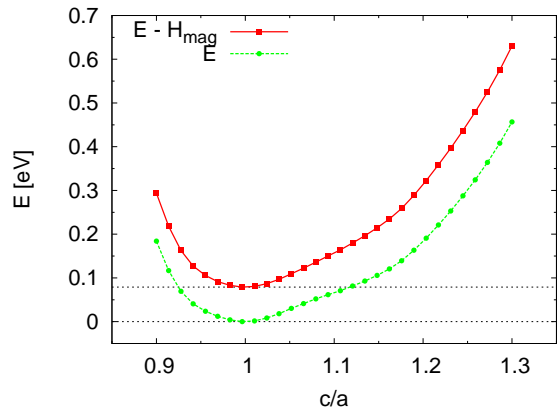}
\caption{\label{fig:emag.u} (Color online) Comparison of the total ground state energy and the magnetic energy
per unit cell in GGA+U.} 
\end{figure}
Since we have computed the parameters $J_v$ and $J_b$, the magnetic Heisenberg energy in Eq.(\ref{Hmag}) can be
determined as a function of $c/a$ (in the fully ferromagnetic case with $S_{+}=S_{-}$).
Doing so, we have a rough estimate of the magnetic energy in the system. To understand its contribution to the
total energy, we subtract ${\cal H}_{\rm mag}$ from the total energy at each $c/a$ value. The result is shown in 
Figs.~\ref{fig:emag.gga} and~\ref{fig:emag.u} for GGA and GGA+U respectively. 
As can be seen from Fig.~\ref{fig:emag.gga}, when the magnetic
energy is subtracted from the total energy, the minimum at $c/a \simeq 1.2$ disappears in GGA. In GGA+U,
the subtracted magnetic energy does not change the overall profile of the dependence on $c/a$. Notice that the 
magnetic energy is larger in magnitude in GGA than in GGA+U. Although the magnetization of Mn atoms is higher
in GGA+U, the effective exchange parameters are significantly
smaller in magnitude (as seen in Fig.~\ref{fig:jall}), leading to a smaller absolute magnetic energy. There is always
an upper bound in Mn magnetization, which cannot be exceeded no matter
how large Hubbard $U$ is. However, larger Hubbard
$U$ will always lower $\vert J \vert$ (in consistence with the perturbation expansion $J \sim t^2/U$), thus leading to a lower 
absolute magnetic energy. 
Therefore, we argue that the appearance of the stable structure at $c/a \simeq 1.2$ is determined by magnetic energy.
Then, the magnitude of the exchange interactions $J_v$ and $J_b$ needs to be determined accurately for 
a correct description of the structural properties. In fact, the overestimation of the exchange parameters
leads to the (spurious) structural minimum at $c/a \simeq 1.2$ for the stoichiometric compound with GGA,
which is eliminated with the Hubbard correction. 

%
\section{\label{sec:mag}Magnetic Coupling Mechanisms}
Due to the large separation between Mn atoms in the crystal, it is very unlikely that a direct exchange interaction
between Mn atoms could be responsible for the magnetic interactions. Instead, Ni and Ga atoms between Mn atoms 
are more likely to mediate super-exchange~\cite{anderson-sxc} type interactions. Indeed, 
hints of this possibility were highlighted in the previous section. For instance, 
we had found that the oscillations in the effective exchange parameters between Mn atoms
show strong similarity to the oscillations observed in the Ni $d$ state magnetization.
Magnetic interactions mediated by conduction electrons (like Ni $d$ and Ga $p$ states) have long been 
known, characterized by an effective exchange parameter having an oscillatory behavior, 
known as Ruderman-Kittel-Kasuya-Yosida (RKKY) interaction~\cite{rkky}. The RKKY interaction arises from the
polarization of free electrons in response to the presence of a magnetic impurity which, reaching other 
impurities, results in a magnetic interaction between them. 
The oscillatory behavior in the magnetic interaction strength can be interpreted as result of electronic screening. 
The free electron response to a perturbation is measured by the Lindhard function, 
which has an oscillatory behavior in real space and decays as the third power of distance
(i.e. $\sim \cos(k_F\, R)/(k_F R)^3$, where $k_F$ is the Fermi momentum) between the impurities. 
If localized Mn $d$ electrons are treated as a periodic array of ``magnetic impurities'', and the Ni $d$ and Ga $p$ 
electrons are treated as conduction electrons, then a magnetic interaction of RKKY type can be expected. Indeed,
this possibility has been considered in the literature before
~\cite{khoi-rkky,heusler-rkky1,heusler-rkky2,heusler-rkky3,heusler-rkky4}.

Before discussing the relevance of RKKY interactions for our system, we first provide a brief summary of the results
from a model Hamiltonian approach, which brings some insight into the problem. The RKKY interaction can be
derived starting from the Anderson impurity model~\cite{anderson-imp,anderson-imp2}, which can be
expressed as
\begin{eqnarray}
{\cal H} &=& \sum_{{\bf k}, \sigma} \epsilon_{{\bf k}}\, {\hat n}_{{\bf k}\, \sigma} + 
 \sum_{d} E\, \left( {\hat n}_{d\, \uparrow} + {\hat n}_{d\, \downarrow} \right) +
 \sum_{d} U\, {\hat n}_{d\, \uparrow}\, {\hat n}_{d\, \downarrow} \nonumber\\ 
 && \qquad +
 \sum_{{\bf k}, d, \sigma} \left( V_{d\, {\bf k}}\, {\hat c}_{{\bf k} \sigma}^{\dag}\, {\hat c}_{d\, \sigma} 
        + {\rm h.c.} \right) \label{ham:anderson} 
\end{eqnarray}
where $\epsilon_{{\bf k}}$ denote the energy levels and
${\hat n}_{{\bf k}\, \sigma} = {\hat c}_{{\bf k}\, \sigma}^{\dag}\, {\hat c}_{{\bf k}\, \sigma}$ is the 
number operator for free electrons respectively (or Bloch electrons in the context of a periodic system).
$E$ denotes the energy of electrons localized on the impurity,
$U$ is the Coulomb repulsion between two electrons localized on the same impurity and 
${\hat n}_{d\, \uparrow (\downarrow)}$ are the number operators. $V_{d\, {\bf k}}$ is the hopping amplitude
of electrons from the impurity site to the free electron states, and so it accounts for interactions. 
In writing the Hamiltonian (\ref{ham:anderson}), we have ignored the direct coupling between impurity sites:
$\sum_{d,d'} ( V_{d d'}\, {\hat c}_{d \sigma}^{\dag} {\hat c}_{d' \sigma} + {\rm h.c.} )$. In fact, 
we assume that such interactions are negligible, due to spatial separation between Mn sites.
Moreover, it was also shown in Ref.~\onlinecite{anderson-imp2} that such interactions are generally 
anti-ferromagnetic for two impurities, when the Fermi level well separates the impurity energies $E$ and $E+U$.
This is the case for our system as can be seen from Figs.~\ref{fig:pdos.gga} and~\ref{fig:pdos.u}.
For both GGA and GGA+U, the filled and empty $d$-bands of Mn are well separated.
The simplest realization of Eq. (\ref{ham:anderson}) is when there is only a single impurity ($d=1$). 
In this case, one can derive an effective exchange interaction parameter between the impurity 
states and conduction electrons in second order perturbation theory as~\cite{SW}
\begin{equation}
J_{d\, {\bf k}} \simeq \frac{2\, \vert V_{d\, {\bf k}} \vert^2\, U}{\vert E \vert\, 
 \left( U - \vert E \vert \right) } \label{Jdk}
\end{equation}
where it is assumed that the impurities are maximally polarized $\langle n_{d\, \uparrow} \rangle \sim 1$,
$\langle n_{d\, \downarrow} \rangle \sim 0$,  $E < 0$, $E + U > 0$ and the Fermi energy is set to zero.
Similarly, the magnetization of free electrons, as a response to the magnetization of the impurity, can be
derived as~\cite{anderson-imp}
\begin{equation}
\mu_{{\bf k}} \simeq \frac{1}{2}\, \vert V_{d\, {\bf k}} \vert^2\, \frac{d\rho}{d\epsilon}\,
       {\rm ln}\left[ \frac{ E^2 + \Delta^2 }{ (E+U)^2 + \Delta^2} \right] \label{mu_free}
\end{equation}
where it is assumed that the density of free electron states, $\rho(\epsilon)$, is a slowly varying function
of $\epsilon$. In the above
equation, $\Delta = \pi\, \langle \vert V_{d\, {\bf k}} \vert^2 \rangle \, \rho(\epsilon)$
is a measure of the mixing between free electron and impurity states (More precisely, $\Delta$ corresponds to 
the line width of impurity states, due to interactions with the free electrons). Comparing equations
(\ref{Jdk}) and (\ref{mu_free}) we see that the exchange interaction between the impurity and conduction
electrons are proportional to the same factor $\vert V_{d\, {\bf k}} \vert^2$, so the larger the 
$\vert V_{d\, {\bf k}} \vert^2$,
the larger the magnetization of free electrons would be. Moreover, the interaction between 
the conduction electrons and the impurity states is anti-ferromagnetic ($J_{d\, {\bf k}} > 0$). The
RKKY interaction can be obtained from a fourth order perturbation theory starting from a two impurity 
Hamiltonian analogous to Eq.(\ref{ham:anderson}). The resulting impurity-impurity interaction energy is
given by~\cite{rkky}
\begin{equation}
{\cal H}_{d\, d} = -J_{d\, d}\, f( k_F\, R)\, {\bf S}_{1} \cdot {\bf S}_{2} \label{hrkky}
\end{equation}
where
\begin{equation} 
J_{d\, d} \sim m\, k_F^4\, \vert J_{d\, {\bf k}} \vert^2 \,\,\,\, , \,\,\,\,
f(x) = \frac{ 2\, x\, \cos(2x) - \sin(2x) }{x^4} \label{Jdd}
\end{equation}
where $m$ is the electron mass (replaced with effective mass $m^*$ in a crystal), $R$ is the distance between
impurities and ${\bf S}_{1,2}$ are their spins. For $k_F\, R \ll 1$, the interaction between the
impurities is ferromagnetic, and its strength decays with the third power of their distance. In Ni$_2$MnGa, we 
treat the Ni $d$ and Ga $p$ states as the conduction electrons. Notice that Ga $p$ states
are polarized anti-ferromagnetically with respect to Mn $d$ states, which is in agreement with the picture 
Anderson model provides. On the other hand, Ni $d$ states are ferromagnetically ordered. This is probably related
to the fact that the above picture does not take into account the topology of the Fermi surface, which is
more important for Ni $d$ states. Indeed, we observe from Eq.(\ref{mu_free}) that the sign of the free electron
polarization can change depending on the derivative of the density of states.

In an alloy, the situation is more complicated than a model with impurities embedded in free electrons
for mainly two reasons; first, the ``impurities'' i.e. localized
Mn $d$ states form a lattice and second, the conduction electrons that mediate the magnetic interactions are
not free and therefore the strength of the interaction depends on the nontrivial topology of the Fermi surface.
However, several studies in the literature have found (using a frozen magnon approach)
the oscillatory behaviour of the Mn-Mn exchange
parameter indicated in Eq.(\ref{Jdd})
~\cite{heusler-rkky1,heusler-rkky2,heusler-rkky3,heusler-rkky4}.
Indeed, the possibility that the magnetic interactions are mediated by ``X'' and ``Z'' atoms 
of generic X$_2$YZ Heusler alloys was considered long before in the literature~\cite{heusler-sx},
by studying the features in the density of states. In a similar way, we will use some basic arguments
to understand the nature of the magnetic interactions in our system, in light of the Anderson impurity 
model and of RKKY interaction. A more rigorous study would require the calculation of the Lindhard  
susceptibility using a linear response approach~\cite{lee-chi,wilson-chi1,wilson-chi2}. 
Comparing Eqs.(\ref{Jdk}),(\ref{mu_free}) and (\ref{Jdd}), we observe that the strength of the interaction 
between the impurities is proportional to the square of the magnetization of conduction electrons. Using this
result, we model the exchange parameters $J_v$ and $J_b$ in order to investigate the 
relative strength of RKKY interactions mediated by Ni $d$ and Ga $p$ states separately as
\begin{eqnarray}
&& J_b^{\rm model} = \left( j_{1 b}\, \mu_{\rm Ni}^2 + j_{2 b}\, \mu_{Ga}^2 \right)\, r_b^{-3} \nonumber\\
&& J_v^{\rm model} = \left( j_{1 v}\, \mu_{\rm Ni}^2 + j_{2 v}\, \mu_{Ga}^2 \right)\, r_v^{-3} \label{jfit}
\end{eqnarray}
where $r_b$ and $r_v$ are the relevant distances between Mn atoms (more precisely, their ratio to the
cubic lattice parameter) and
the coefficients $j_{1,2\, b}$ and $j_{1,2\, v}$ are constants to be determined by fitting to the 
data obtained from GGA and GGA+U calculations in the magnetic energy given in Eq.(\ref{Hmag}). 
\begin{figure}[!ht]
\includegraphics[width=0.45\textwidth]{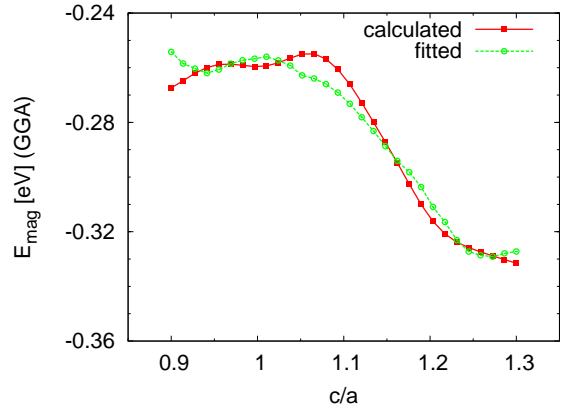}
\caption{\label{fig:fit.gga} (Color online) Comparison of the computed and fitted magnetic Heisenberg energies
in GGA.}
\end{figure}
\begin{figure}[!ht]
\includegraphics[width=0.45\textwidth]{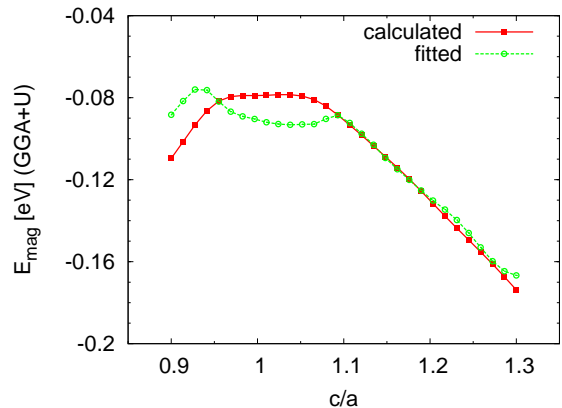}
\caption{\label{fig:fit.u} (Color online) Comparison of the computed and fitted magnetic Heisenberg energies
in GGA+U.}
\end{figure}
The functional form of Eqs.(\ref{jfit})
does not take into account effects coming from the Fermi surface topology that determines the 
($c/a$-dependent) screening
by conduction electrons (especially from Ni $d$ states). Instead,
they provide an average dependence on the magnetic moments.
In order to find the parameters $j_{1,2\, b}$ and $j_{1,2\, v}$, we fit the expressions in Eq.(\ref{jfit})
using Eq.(\ref{Hmag}) ($E_{\rm model}$) to the Heisenberg energy obtained from constrained DFT calculations
for every $c/a$. 
The fit is obtained by the least squares method and the values obtained for $j_{1,2\, b}$ and $j_{1,2\, v}$ 
are reported in Table.~\ref{tab:fit}. The comparison of the fitted
and the computed magnetic energies is shown in Figs.~\ref{fig:fit.gga} and~\ref{fig:fit.u}.
\begin{table}[b]
\caption{\label{tab:fit}
Fitted parameters in the model for $J_v$ and $J_b$ found in GGA and GGA+U.}
\begin{ruledtabular}
\begin{tabular}{ccc}
& GGA $[{\rm Ry}/{\mu}_B^2]$ & GGA+U $[{\rm Ry}/{\mu}_B^2]$  \\
\hline
$j_{1b}$  &  $-1.94 \times 10^{-4}$  &  $-2.73 \times 10^{-4}$ \\
$j_{2b}$  &  $5.63 \times 10^{-4}$  &  $-1.61 \times 10^{-4}$ \\
$j_{1v}$  &  $4.99 \times 10^{-5}$  &  $1.68 \times 19^{-4}$ \\
$j_{2v}$  &  $-1.19 \times 10^{-3}$  &  $-3.17 \times 10^{-4}$
\end{tabular}
\end{ruledtabular}
\end{table}
The quality of the fits can be tested quantitatively by evaluating the ratio 
${\tilde\chi}^2 = \sum_i \left[ E_{\rm mag}(i) - E_{\rm model}(i) \right]^2/\sum_i E_{\rm mag}^2(i)$,
where $i$ runs over the $c/a$ values considered. For an accurate fit, ${\tilde\chi}^2 \ll 1$.
We find that ${\tilde\chi}^2 \simeq 1.1 \times 10^{-5}$ for GGA and ${\tilde\chi}^2 \simeq 2.5 \times 10^{-4}$
for GGA+U. As can be seen from Table.~\ref{tab:fit}, GGA+U predicts larger exchange parameters (in magnitude)
for the Mn-Ni interaction ($j_{1\, v,b}$) than from GGA, and smaller ones for the Mn-Ga interaction
($j_{2\, v,b}$). The larger Mn-Ga interaction in GGA can actually be understood using Eq.(\ref{Jdk}) and
the density of states. From Fig.~\ref{fig:pdos.gga}, we observe that the lowest energy filled Mn $d$ states 
are peaked around $3.6\, {\rm eV}$ below $E_f$, while the unoccupied Mn $d$ states are peaked around
$1.6\, eV$ above $E_f$. This corresponds to $\vert E \vert \simeq 3.6\, eV$ and 
$-\vert E \vert + U_{\rm eff} \simeq 1.6\, {\rm eV}$ (where $U_{\rm eff}$ is the effective Coulomb interaction
on Mn site) and therefore from Eq.(\ref{Jdk}) we find $J_{Mn-Ga} \simeq 1.8\, \vert V_{Mn-Ga} \vert^2$ in GGA.
Instead, from Fig.~\ref{fig:pdos.u}, we see that the occupied Mn $d$ states are peaked around $-6.4\, {\rm eV}$,
while the unoccupied Mn $d$ states are peaked around $4\, {\rm eV}$, which leads to $\vert E \vert \simeq 6.4\, eV$
and $-\vert E \vert + U_{\rm eff} \simeq 4\, {\rm eV}$. Therefore, $J_{Mn-Ga} \simeq 0.81\, \vert V_{Mn-Ga} \vert^2$
in GGA+U. This value is further lowered by the fact that $\vert V_{Mn-Ga} \vert$ in GGA+U is smaller due to 
stronger localization of electrons on Mn sites. 
Instead, the larger exchange parameters for Mn-Ni interaction in GGA+U is related to the factor
$k_F^4\, f(k_F\, R)$ in Eq.(\ref{Jdd}). Since Ni $d$ states have the dominant contribution
at the Fermi level, the dependence on the Fermi momentum (to be more precise, the Fermi surface) is more important
for the Mn-Ni exchange interaction. However Eq.(\ref{Jdk}) is not sufficient to account for the topology of the Fermi 
surface, and therefore we can not use it to estimate the relative strength of Mn-Ni exchange parameters,
unlike we did for Mn-Ga exchange parameter. Instead, we interpret Mn-Ni exchange parameters $j_{1\, v,b}$ as
describing the magnetic interaction in an average way. 
The inaccuracy to model the Fermi surface topology in the magnetic interactions, 
also explain why the fitted magnetic energies shown in Figs~\ref{fig:fit.gga} and~\ref{fig:fit.u} 
are less accurate for $c/a \la 1.1$ but 
more accurate for $c/a \ga 1.1$. As can be seen from Fig.~\ref{fig:magNi}, the magnetic energy is correlated
with the Ga magnetization for $c/a \ga 1.1$. However for $c/a \la 1.1$, Ga magnetization is smaller and Ni $d$ states
start to have larger influence on the exchange parameters 
and due to the dependence on the Fermi surface topology,
the inaccuracy in determining $J_{Mn-Ni}$ becomes more relevant. Another trend we observe is that 
the exchange parameters $\vert j_{1\,v,b} \vert$ are in general smaller than $\vert j_{2\, v,b}\vert$. 
The only exception is $\vert j_{2b} \vert < \vert j_{1b} \vert$ in GGA+U.
These results indicate that
the stable structure of Ni$_2$MnGa is determined by a competition between the exchange parameters of Mn-Ni and
Mn-Ga interactions, that determines the magnetic energy as a function of $c/a$. The model in Eq.(\ref{jfit})
provides a simple average measure of the dependence of magnetic interactions on the magnetization of Ga and
Ni atoms, which can be used to explain different mechanisms that determines the relative stability of cubic
and tetragonal phases. Specifically, Eq.(\ref{jfit}) provides an accurate estimate of the 
dependence of the magnetic energy on $c/a$ and shows that Mn-Ni super-exchange is prevalent for $c/a \la 1$,
while Mn-Ga super-exchange is prevalent for $c/a \ga 1$. 
In Table.~\ref{tab:fit}, we report the effective interaction strengths related to
these mechanisms. While the Mn-Ga exchange interaction is easily realized through the Anderson impurity model, 
the Mn-Ni interaction requires a careful study of the Lindhard electronic response function, that is closely 
related to the Fermi surface topology. Indeed, the softening of phonon
modes, the dependence of the stability of the structure on the number of valence electrons per unit cell, 
and the appearance of modulated structures have been argued to be connected to the Fermi surface topology
~\cite{opeil-pre,rabe-phon,zheludev-fs}. Our study is a further confirmation of these hypothesis.

\section{\label{sec:extension}Electron injection and Off-Stoichiometric Alloys}
\subsection{\label{sub:electron} Extra electrons in the unit cell}
\begin{figure}[!ht]
\includegraphics[width=0.45\textwidth]{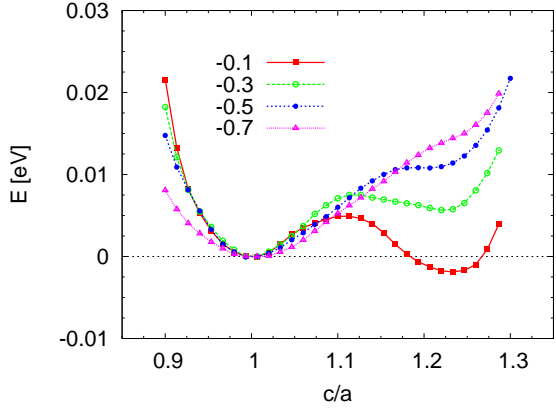}
\caption{\label{fig:evscoa.charge} (Color online) Energy of the unit cell as a function of $c/a$ for 
different values of added charge, using GGA.}
\end{figure}

\begin{figure}[!ht]
\includegraphics[width=0.45\textwidth]{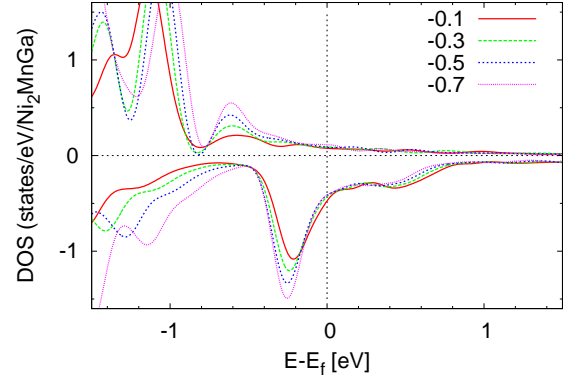}
\caption{\label{fig:nieg} (Color online) Density of states of Ni $e$ states around the Fermi level.
The zero of energy for each curve is set to the Fermi energy of the corresponding charged state,
which increases with charge $e$.}
\end{figure}

\begin{figure}[!ht]
\includegraphics[width=0.45\textwidth]{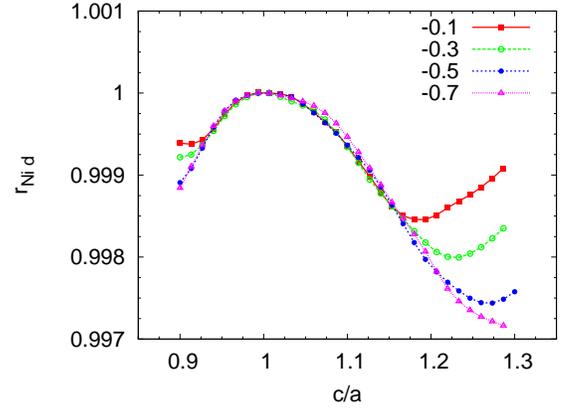}
\caption{\label{fig:rNi.charge} (Color online) $r_{\rm Ni d} \equiv n_{\rm Ni d}/n_{\rm Ni d}^{\rm cubic}$
as a function of $c/a$ for a couple of extra charge.}
\end{figure}

\begin{figure}[!ht]
\includegraphics[width=0.45\textwidth]{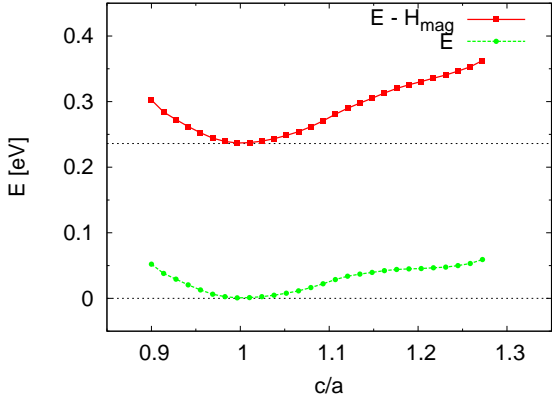}
\caption{\label{fig:emag.-0.5.gga} (Color online) Comparison of the total ground state energy and the magnetic energy
per unit cell in GGA with extra charge $e=0.5$.}
\end{figure}

\begin{figure}[!ht]
\includegraphics[width=0.45\textwidth]{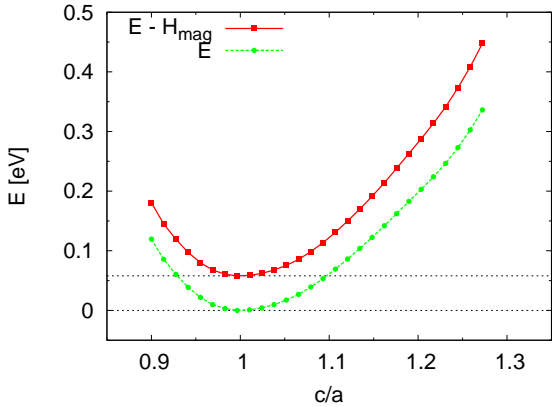}
\caption{\label{fig:emag.-0.5.u} (Color online) Comparison of the total ground state energy and the magnetic energy
per unit cell in GGA+U with extra charge $e=0.5$.}
\end{figure}

The correlation between the total number of valence electrons per atom ($e/a$) in the unit cell and
structural properties of Heusler alloys, such as the Martensitic transition temperature, the stability of
the cubic structure and the vibrational spectra, has been studied
extensively in the literature~\cite{rabe-eoa,rabe-vib,hu-eoa,dai-eoa,lanska-coa}.
In order to investigate the effects of $e/a$, we have introduced fractional electrons in the unit cell
in the range $0.1 \le e \le 0.7$, compensating the extra charge with a uniform positively
charged jellium. In each case, we have relaxed the unit cell and optimized the cubic cell parameter which is
$a_0 \simeq 5.91$ \AA $\,$ for $e = 0.1$ (slightly larger than the original cell) and increases monotonically
up to $a_0 \simeq 6.48$ \AA $\,$ for $e=0.7$. In Fig.~\ref{fig:evscoa.charge} we show 
the ground state energy (the zero of the energy
corresponds to the cubic phase in each case) as a function of $c/a$ for a selection of extra fractional 
charge using the GGA functional.
Although GGA is not sufficient to predict the relative stability of different
structural phases,
it will be the functional of choice in this part to discriminate the effects
of varying $e/a$ on the exchange couplings from those due to the 
electronic localization. The Hubbard (+$U$) correction will be taken into
account when computing the value of the exchange parameters in presence of
a fixed amount of extra electrons.
At the level of the Anderson 
model, GGA overestimates $\vert V_{d {\bf k}} \vert$ in Eq.(\ref{Jdk}) and underestimates the effective on-site
Coulomb repulsion $U$. Instead, by changing $e/a$ we modify the Fermi surface directly and
$U$ is unaffected at first approximation (of course, since the problem is a self consistent one, 
the Fermi surface would eventually affect it, but only at a higher order). 
Indeed, we have investigated the density of states for ground states with extra charge, and found no 
significant difference with the original system, except a shift of the Fermi energy to higher values
for larger $e/a$. The effect of adding charge is most dominant in Ni $e$ states
whose density of states is shown in Fig~\ref{fig:nieg}. 
As can be seen, for larger $e/a$ the peaks 
just below the Fermi level becomes stronger since more electrons need to be accommodated 
below the Fermi energy.
Therefore, the addition of extra charge in the system
modifies the $k_F^4\, f(k_F\, R)$ term in the RKKY interaction of Eq.(\ref{Jdd})
for the free electron case, and an analogous
averaged quantity depending on the Fermi surface corresponding, in our case, to model parameters 
$j_{1\, v,b}$ in Eq.(\ref{jfit}).
Indeed, as we show in Fig.~\ref{fig:rNi.charge},
as the extra charge injected into the system increases, the Friedel-like oscillations (which also 
depend on the same factor $k_F^4\, f(k_F\, R)$ for free electrons) have larger wavelength.
Since the period of charge oscillations is inversely proportional to $k_F$, 
Fig.~\ref{fig:rNi.charge} suggests that the 
larger the charge $e$, the smaller $k_F$ is. Therefore, from Eq.(\ref{Jdd}), we expect a smaller exchange
coupling, that leads to a smaller Heisenberg magnetic energy. This is indeed visible in the total energy 
calculations in Fig.~\ref{fig:evscoa.charge} as the stable structure at $c/a \simeq 1.2$ 
disappears when $e>0.1$. 
In order to verify this claim, we have computed the exchange couplings and the magnetic energy,
as we did in the previous section, for $e=0.5$.  
\begin{figure}[!ht]
\includegraphics[width=0.45\textwidth]{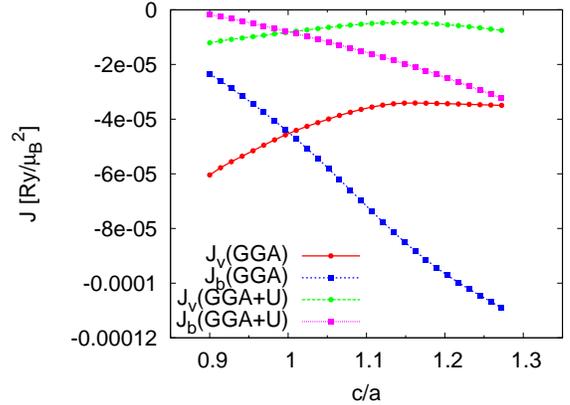}
\caption{\label{fig:jall.-0.5} (Color online) Exchange parameters $J_v$ and $J_b$ computed using GGA and GGA+U
functionals with additional $e=0.5$ electrons.}
\end{figure}
In Fig.~\ref{fig:jall.-0.5}, we show the Mn-Mn exchange parameters $J_v$ and $J_b$ computed with 
constrained DFT calculations using GGA and GGA+U 
functionals ($U$ has the same value as in the original structure). 
Notice that, as was the case for the original neutral cell, the exchange parameters in GGA+U are smaller
than in GGA as expected. Comparing with the values
obtained from the original structure with no extra electrons, shown in Fig.~\ref{fig:jall}, we observe that
the exchange parameters are smaller for $e=0.5$. This can also be seen from 
Figs.~\ref{fig:emag.-0.5.gga} and~\ref{fig:emag.-0.5.u} in comparison with Figs.~\ref{fig:emag.gga} 
and~\ref{fig:emag.u}, where we show the difference between the magnetic 
Heisenberg energy and the total ground state energy.
Moreover, the oscillatory behaviour in the original 
$e=0$ system is also suppressed. Both these observations are in agreement with the previous discussion
based on Friedel-type oscillations having higher period for $e=0.5$, thus leading to 
smaller exchange parameters $J_{v,b}$ with suppressed oscillatory behaviors. Therefore, we expect the 
model given in Eq.(\ref{jfit}) for the exchange parameters to be more precise in case of $e=0.5$, since
the effects of Fermi surface is less pronounced.  
\begin{figure}[!ht]
\includegraphics[width=0.45\textwidth]{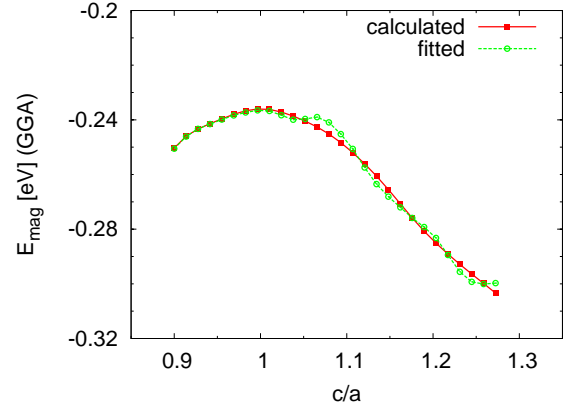}
\caption{\label{fig:fit.-0.5.gga} (Color online) Comparison of the computed and fitted magnetic Heisenberg energies
per unit cell in GGA with $e=0.5$.}
\end{figure}
\begin{figure}[!ht]
\includegraphics[width=0.45\textwidth]{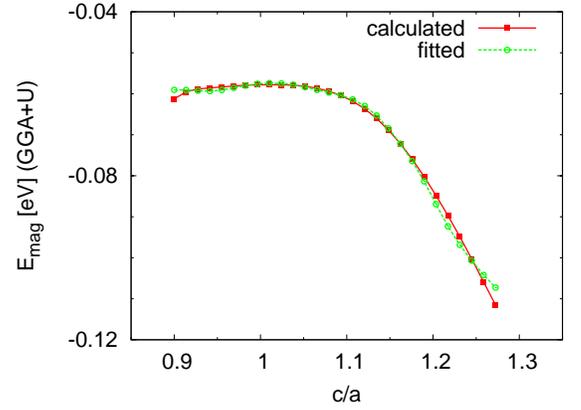}
\caption{\label{fig:fit.-0.5.u} (Color online) Comparison of the computed and fitted magnetic Heisenberg energies
per unit cell in GGA+U with $e=0.5$.}
\end{figure}
Indeed, this is what we observe in Figs.~\ref{fig:fit.-0.5.gga} and~\ref{fig:fit.-0.5.u}. 
In the previous section, we have argued that
the inaccuracies between computed and fitted exchange parameters
arise due to the model in Eq.(\ref{jfit}) being unable to capture effects of 
the Fermi surface topology. In the current case, these effects are suppressed, thus the model adheres more closely 
to the direct calculation. We find that ${\tilde \chi}^2 \simeq 1.92 \times 10^{-6}$ in GGA and 
${\tilde \chi}^2 \simeq 1.2 \times 10^{-5}$ in GGA+U, almost an order of magnitude smaller than 
the case for the neutral cell.
\begin{table}[b]
\caption{\label{tab:fit.-0.5}
Fitted parameters in the model for $J_v$ and $J_b$ found in GGA and GGA+U with extra charge $e=0.5$.}
\begin{ruledtabular}
\begin{tabular}{ccc}
& GGA $[{\rm Ry}/{\mu}_B^2]$ & GGA+U $[{\rm Ry}/{\mu}_B^2]$  \\
\hline
$j_{1b}$  &  $5.94 \times 10^{-5}$  &  $-1.64 \times 10^{-4}$ \\
$j_{2b}$  &  $-9.47 \times 10^{-4}$  &  $9.67 \times 10^{-5}$ \\
$j_{1v}$  &  $-6.66 \times 10^{-5}$  &  $5.97 \times 19^{-5}$ \\
$j_{2v}$  &  $-2.67 \times 10^{-4}$  &  $-2.30 \times 10^{-4}$
\end{tabular}
\end{ruledtabular}
\end{table}
In Table~\ref{tab:fit.-0.5} we report the values of the fitted parameters for $J_v$ and $J_b$, which 
are on average smaller than the fitted values of Table~\ref{tab:fit} for the neutral cell as expected.
As was the case for the neutral cell, the Mn-Ga interaction is 
suppressed in GGA+U compared to GGA. For Mn-Ni interactions, we again observe the fitting parameters
to be an average of the oscillatory behavior, with random signs. Notice that
the stable structure is again determined from a competition between Mn-Ni and Mn-Ga super-exchange
interactions, as described on average by Eq.(\ref{jfit}).

We would like to remark that 
the suppression of the tetragonal structure in GGA+U, as demonstrated in 
section~\ref{sec:GGA} does not preclude the stabilization of the martensitic phase
for the stoichiometric compound. As we have shown, the relative stability of the cubic and non-modulated 
tetragonal phases for the stoichiometric compound is determined by magnetic energy. Instead, 
the appearance of modulated structures, as well as phonon mode softening, are expected to be related 
to Fermi surface nesting~\cite{opeil-pre,rabe-phon,zheludev-fs}. Magnetic interactions also 
depend on Fermi surface properties, as discussed in this section. However, the dominant
contributions come from electronic localization on Mn $d$ states, as demonstrated in the
previous section. Since the Hubbard $U$ correction does not
modify the states responsible for the Fermi surface nesting (Ni $e$ states), GGA+U is not 
expected to suppress modulated structures. Indeed, preliminary GGA+U calculations 
(not reported here) on the 5M modulated
structure revealed the existence of stable martensitic phase with $c/a \simeq 0.92$, in agreement
with experiments~\cite{kokorin-modul,martynov-modulated,dai-eoa}. Moreover,
larger $e/a$ values suppress the tetragonal phase as can be seen in 
Fig.~\ref{fig:evscoa.charge}, in analogy with the effect of $U$. At the same time it has been shown 
in the literature that larger values of
$e/a$ lead to stronger softening of phonon modes and some of the elastic constants
~\cite{rabe-eoa,hu-eoa}. These results, which are both based on GGA, are 
indications of the fact that the two phenomena (pure tetragonal distortions and modulations/phonon mode softening) 
are largely independent.
\subsection{Off-Stoichiometric Compounds}
The effects of extra Mn in Ga sites has been discussed in detail in the literature before
~\cite{hu-eoa,lanska-coa,enkovaara-coa,hu-x,hu-x2} and Mn in place of Sn, for the related Heusler
alloy Ni$_2$MnSn in Ref.~\onlinecite{Ni2MnSn}. It has been shown that the Mn in Ga sites prefers an 
anti-ferromagnetic ordering with respect to normal Mn sites~\cite{enkovaara-coa,Ni2MnSn}. Moreover,
the samples that undergo a martensitic transition to a tetragonal structure with $c/a \simeq 1.2$ always
have larger Mn content than the stoichiometric compounds in Ni$_2$MnGa~\cite{sozinov-aniso,lanska-coa,magmech-5,dai-eoa}. 
In the picture we have provided in the previous sections, the structure with $c/a \simeq 1.2$ is stabilized 
by magnetic interactions between Mn atoms. In this section, we show that the 
$c/a \simeq 1.2$ phase appears only in Ni$_2$Mn$_{1+x}$Ga$_{1-x}$ compounds, 
when $x \ga 0.5$, using the GGA+U functional. Instead, the GGA functional produces inaccurate 
stability profile as a function of $c/a$ for these alloys. 

\begin{figure}[!ht]
\includegraphics[width=0.45\textwidth]{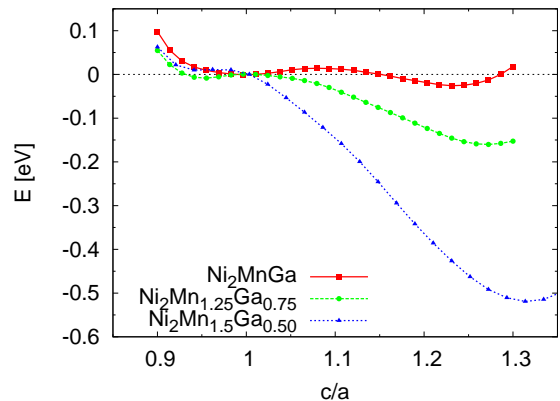}
\caption{\label{fig:evscoa.gga.x} (Color online) Energy of the ground state as a function of $c/a$ at constant volume
per unit cell in GGA. The zero of energy is set to be at $c/a=1$ in each case.}
\end{figure}

\begin{figure}[!ht]
\includegraphics[width=0.45\textwidth]{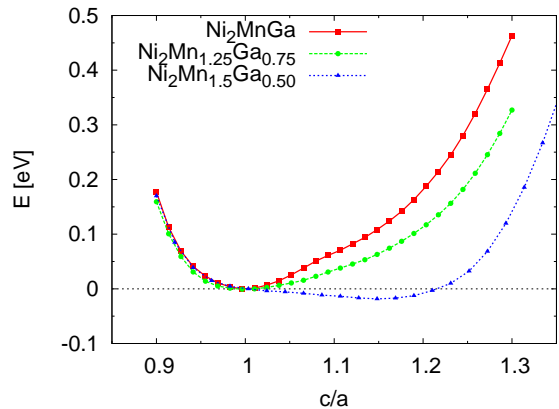}
\caption{\label{fig:evscoa.u.x} (Color online) Energy of the ground state as a function of $c/a$ at constant volume
per unit cell in GGA+U. The zero of energy is set to be at $c/a=1$ in each case.}
\end{figure}

The total energy vs $c/a$ profiles for Ni$_2$MnGa, Ni$_2$Mn$_{1.25}$Ga$_{0.75}$ and Ni$_2$Mn$_{1.5}$Ga$_{0.5}$ using
GGA and GGA+U are shown in Figs.~\ref{fig:evscoa.gga.x} and~\ref{fig:evscoa.u.x}. For the off-stoichiometric
compounds, we have used a 16 atom supercell, and replaced one Ga with a Mn in case of Ni$_2$Mn$_{1.25}$Ga$_{0.75}$,
and two Ga with Mn atoms in case of Ni$_2$Mn$_{1.25}$Ga$_{0.75}$. In the latter case, the second Mn impurity is
placed in the same $x-y$ plane with the first Mn impurity. We have also studied the system where the second 
Mn impurity is placed in a different $x-y$ plane with respect to the first Mn impurity and found that both systems have 
similar electronic and structural properties. For both off-stoichiometric compounds, we have relaxed the unit cell,
optimized the cubic lattice parameter (using GGA) and studied constant volume tetragonal distortions.
We have used the same $U$ on Mn atoms calculated for the 
stoichiometric cell. Although a more precise approach should involve a re-calculation of the Hubbard $U$ in this case, 
as a first approximation, the same $U$ value is sufficient to study the general trends.
We have found that an antiferromagnetic ordering of extra Mn atoms with respect to the original Mn atoms
is energetically favored in agreement with the previous studies~\cite{enkovaara-coa,Ni2MnSn}. This is expected, since
the extra Mn impurities are nearest neighbors to the original Mn atoms
(instead of being separated by Ni and Ga atoms),
and therefore the exchange interaction 
is mainly mediated by a direct hopping of electrons between them (i.e. through the term 
$\sum_{d,d'} ( V_{d d'}\, {\hat c}_{d \sigma}^{\dag} {\hat c}_{d' \sigma} + {\rm h.c.} )$, we have
neglected previously). This type of direct hopping results in an anti-ferromagnetic coupling 
with an exchange parameter $J \sim \vert V_{dd'} \vert^2/U$. 
Addition of extra Mn notably increases the magnitude of magnetic Heisenberg 
energy, leading to a stronger decrease of total energy in both GGA and GGA+U. However, since GGA
overestimates the magnetic couplings, it yields a local energy minimum at $c/a < 1$ 
and a global energy minimum at $c/a > 1.2$ for Ni$_2$Mn$_{1.25}$Ga$_{0.75}$. This result is 
in contradiction with experimental data (an energy minimum for $c/a < 1$ was observed only for   
modulated structures). In the case of Ni$_2$Mn$_{1.5}$Ga$_{0.5}$, there is no local minimum for $c/a < 1.2$ and 
the global minimum is placed at $c/a \simeq 1.3$ overestimating the experimental value. Instead, 
the GGA+U functional yields only a global minimum for $c/a \simeq 1.15$ for Ni$_2$Mn$_{1.5}$Ga$_{0.5}$. 
This finding is also in agreement with the case of Ni$_2$Mn$_{1+x}$Sn$_{1-x}$ where the Martensitic
transition (to a non-modulated structure) is observed only when $x \ga 0.5$~\cite{Ni2MnSn}. 

\section{\label{sec:conclusion}Conclusions}

In this work we have studied the electronic and structural properties of Ni$_2$MnGa and investigated the 
relative stability of cubic austenite and non-modulated tetragonal martensite phases. We have shown that 
the relative stability of these phases critically depends on the magnetic energy due to exchange 
interactions between Mn atoms.
This result was demonstrated by modelling the magnetic energy with the Heisenberg Hamiltonian and computing
the effective exchange couplings with constrained DFT calculations for every structure.
Comparing the results obtained from GGA and GGA+U functionals, we showed that electronic localization 
has an important effect on the magnetic energy and on the relative stability of the two structures.
In particular, GGA+U removes the spurious energy minimum at $c/a \simeq 1.2$ for the stoichiometric compound. 
An explanation of these results
was obtained by using the Anderson impurity model. Within this model, we have treated
the Mn $d$ orbitals as ``magnetic impurities'' embedded in conduction electrons on Ni $d$ and Ga $p$ orbitals,
which mediate RKKY type interactions between Mn atoms. 
Using this theoretical framework, we were able to provide an explanation for the relative stability of cubic and 
non-modulated tetragonal structures in various circumstances.
We have also formulated a simplified model for the Mn-Mn exchange interaction
based on a linear combination of two contributions: (1) super-exchange mediated by Ga $p$ states and 
(2) super-exchange mediated by Ni $d$ states and were
able to uncover the competition between these mechanisms in determining the stable structure.
This model was also used to study the effect of $e/a$, and it showed 
that the stability of the non-modulated martensitic phase is suppressed by increasing the number of electrons.
We also studied off-stoichiometric compounds with excess Mn (substituting Ga atoms). 
We have shown that, in agreement with experiments, the non-modulated 
martensitic structure with $c/a > 1$ can only be stabilized 
by excess Mn content,
which increases the exchange couplings $J$ between Mn atoms.
This result was obtained using the GGA+U functional. The GGA 
functional predicts, instead, the non modulated tetragonal phase
to be stable at any composition, in contradiction with observations.
 
The study offered in this work shows that a precise account of magnetism and magnetic interactions is 
essential to predict the relative stability of different phases and to rationalize the experimental observations.
The simple RKKY interaction is able to explain most of the physics of the
systems studied in this paper and to capture the effects of doping. 
In spite of the metallic character of these systems,
DFT+U proved to be crucial for an accurate description of the localization
of electrons and of magnetic properties that play a key role 
in the relative stability of different structural phases. 
To the best of our knowledge, DFT+U was
not previously used to study the electronic structure of Ni$_2$MnGa. However, a different 
Heusler alloy Co$_2$Mn$_{1-x}$Fe$_x$Si was studied with DFT+U~\cite{chadov-dftu} and the importance
of electronic correlations was stressed. 
In light of these results, we argue that the study of phonons, Fermi 
surface nesting 
and off-stoichiometric alloys should be revisited using the DFT+U functional.
A recently developed extension of DFPT~\cite{baroni01} to the DFT+U 
scheme~\cite{phononu} will greatly facilitate the development of these
studies.
Our future calculations will use the theoretical framework developed in this work
to explore the effects of various substitutional impurities and 
to characterize modulated martensitic structures.


\begin{acknowledgments}
This work was partially supported by the NSF Grant EAR-0810272. 
We are grateful to the Minnesota Supercomputing Institute for providing 
the computational resources necessary to develop this study.
\end{acknowledgments}

\bibliography{hc_ni2mnga}

\providecommand{\noopsort}[1]{}\providecommand{\singleletter}[1]{#1}%
\begin{thebibliography}{84}
\expandafter\ifx\csname natexlab\endcsname\relax\def\natexlab#1{#1}\fi
\expandafter\ifx\csname bibnamefont\endcsname\relax
  \def\bibnamefont#1{#1}\fi
\expandafter\ifx\csname bibfnamefont\endcsname\relax
  \def\bibfnamefont#1{#1}\fi
\expandafter\ifx\csname citenamefont\endcsname\relax
  \def\citenamefont#1{#1}\fi
\expandafter\ifx\csname url\endcsname\relax
  \def\url#1{\texttt{#1}}\fi
\expandafter\ifx\csname urlprefix\endcsname\relax\def\urlprefix{URL }\fi
\providecommand{\bibinfo}[2]{#2}
\providecommand{\eprint}[2][]{\url{#2}}

\bibitem[{\citenamefont{Brown et~al.}(2007)\citenamefont{Brown, Gandy, Ishida,
  Kainuma, Kanomata, Matsumoto, Morito, Neumann, Oikawa, Ouladdiaf
  et~al.}}]{msm-rev}
\bibinfo{author}{\bibfnamefont{P.}~\bibnamefont{Brown}},
  \bibinfo{author}{\bibfnamefont{A.}~\bibnamefont{Gandy}},
  \bibinfo{author}{\bibfnamefont{K.}~\bibnamefont{Ishida}},
  \bibinfo{author}{\bibfnamefont{R.}~\bibnamefont{Kainuma}},
  \bibinfo{author}{\bibfnamefont{T.}~\bibnamefont{Kanomata}},
  \bibinfo{author}{\bibfnamefont{M.}~\bibnamefont{Matsumoto}},
  \bibinfo{author}{\bibfnamefont{H.}~\bibnamefont{Morito}},
  \bibinfo{author}{\bibfnamefont{K.}~\bibnamefont{Neumann}},
  \bibinfo{author}{\bibfnamefont{K.}~\bibnamefont{Oikawa}},
  \bibinfo{author}{\bibfnamefont{B.}~\bibnamefont{Ouladdiaf}},
  \bibnamefont{et~al.}, \bibinfo{journal}{J. Magn. Magn. Mater.}
  \textbf{\bibinfo{volume}{310}}, \bibinfo{pages}{2755} (\bibinfo{year}{2007}).

\bibitem[{\citenamefont{Krenke et~al.}(2007)\citenamefont{Krenke, Duman, Acet,
  Wassermann, Moya, Ma{\~n}osa, Planes, Suard, and Ouladdiaf}}]{magn-1}
\bibinfo{author}{\bibfnamefont{T.}~\bibnamefont{Krenke}},
  \bibinfo{author}{\bibfnamefont{E.}~\bibnamefont{Duman}},
  \bibinfo{author}{\bibfnamefont{M.}~\bibnamefont{Acet}},
  \bibinfo{author}{\bibfnamefont{E.}~\bibnamefont{Wassermann}},
  \bibinfo{author}{\bibfnamefont{X.}~\bibnamefont{Moya}},
  \bibinfo{author}{\bibfnamefont{L.}~\bibnamefont{Ma{\~n}osa}},
  \bibinfo{author}{\bibfnamefont{A.}~\bibnamefont{Planes}},
  \bibinfo{author}{\bibfnamefont{E.}~\bibnamefont{Suard}}, \bibnamefont{and}
  \bibinfo{author}{\bibfnamefont{B.}~\bibnamefont{Ouladdiaf}},
  \bibinfo{journal}{Phys. Rev. B} \textbf{\bibinfo{volume}{75}},
  \bibinfo{pages}{104414} (\bibinfo{year}{2007}).

\bibitem[{\citenamefont{Inoue et~al.}(2000)\citenamefont{Inoue, Enami,
  Yamaguchi, Ohoyama, Morii, Matsuoka, and Inoue}}]{magn-2}
\bibinfo{author}{\bibfnamefont{K.}~\bibnamefont{Inoue}},
  \bibinfo{author}{\bibfnamefont{K.}~\bibnamefont{Enami}},
  \bibinfo{author}{\bibfnamefont{Y.}~\bibnamefont{Yamaguchi}},
  \bibinfo{author}{\bibfnamefont{K.}~\bibnamefont{Ohoyama}},
  \bibinfo{author}{\bibfnamefont{Y.}~\bibnamefont{Morii}},
  \bibinfo{author}{\bibfnamefont{Y.}~\bibnamefont{Matsuoka}}, \bibnamefont{and}
  \bibinfo{author}{\bibfnamefont{K.}~\bibnamefont{Inoue}}, \bibinfo{journal}{J.
  Phys. Soc. Jpn.} \textbf{\bibinfo{volume}{69}}, \bibinfo{pages}{3485}
  (\bibinfo{year}{2000}).

\bibitem[{\citenamefont{Cherechukin
  et~al.}(2001{\natexlab{a}})\citenamefont{Cherechukin, Dikshtein, Ermakov,
  Glebov, Koledov, Kosolapov, Shavrov, Tulaikova, Krasnoperov, and
  Takagi}}]{magn-3}
\bibinfo{author}{\bibfnamefont{A.}~\bibnamefont{Cherechukin}},
  \bibinfo{author}{\bibfnamefont{I.}~\bibnamefont{Dikshtein}},
  \bibinfo{author}{\bibfnamefont{D.}~\bibnamefont{Ermakov}},
  \bibinfo{author}{\bibfnamefont{A.}~\bibnamefont{Glebov}},
  \bibinfo{author}{\bibfnamefont{V.}~\bibnamefont{Koledov}},
  \bibinfo{author}{\bibfnamefont{D.}~\bibnamefont{Kosolapov}},
  \bibinfo{author}{\bibfnamefont{V.}~\bibnamefont{Shavrov}},
  \bibinfo{author}{\bibfnamefont{A.}~\bibnamefont{Tulaikova}},
  \bibinfo{author}{\bibfnamefont{E.}~\bibnamefont{Krasnoperov}},
  \bibnamefont{and} \bibinfo{author}{\bibfnamefont{T.}~\bibnamefont{Takagi}},
  \bibinfo{journal}{Phys. Lett. A} \textbf{\bibinfo{volume}{291}},
  \bibinfo{pages}{175} (\bibinfo{year}{2001}{\natexlab{a}}).

\bibitem[{\citenamefont{Entel et~al.}(2008)\citenamefont{Entel, Gruner,
  Adeagbo, and Zayak}}]{magn-4}
\bibinfo{author}{\bibfnamefont{P.}~\bibnamefont{Entel}},
  \bibinfo{author}{\bibfnamefont{M.}~\bibnamefont{Gruner}},
  \bibinfo{author}{\bibfnamefont{W.}~\bibnamefont{Adeagbo}}, \bibnamefont{and}
  \bibinfo{author}{\bibfnamefont{A.}~\bibnamefont{Zayak}},
  \bibinfo{journal}{Mater. Sci. Eng., A} \textbf{\bibinfo{volume}{481}},
  \bibinfo{pages}{258} (\bibinfo{year}{2008}).

\bibitem[{\citenamefont{Ma et~al.}(2008{\natexlab{a}})\citenamefont{Ma, Zhang,
  Yu, Zhu, Chen, Wu, Liu, Qu, and Li}}]{magn-5}
\bibinfo{author}{\bibfnamefont{L.}~\bibnamefont{Ma}},
  \bibinfo{author}{\bibfnamefont{H.}~\bibnamefont{Zhang}},
  \bibinfo{author}{\bibfnamefont{S.}~\bibnamefont{Yu}},
  \bibinfo{author}{\bibfnamefont{Z.}~\bibnamefont{Zhu}},
  \bibinfo{author}{\bibfnamefont{J.}~\bibnamefont{Chen}},
  \bibinfo{author}{\bibfnamefont{G.}~\bibnamefont{Wu}},
  \bibinfo{author}{\bibfnamefont{H.}~\bibnamefont{Liu}},
  \bibinfo{author}{\bibfnamefont{J.}~\bibnamefont{Qu}}, \bibnamefont{and}
  \bibinfo{author}{\bibfnamefont{Y.}~\bibnamefont{Li}}, \bibinfo{journal}{Appl.
  Phys. Lett.} \textbf{\bibinfo{volume}{92}}, \bibinfo{pages}{032509}
  (\bibinfo{year}{2008}{\natexlab{a}}).

\bibitem[{\citenamefont{Vasil'ev et~al.}(2003)\citenamefont{Vasil'ev,
  Buchel'nikov, Takagi, Khovailo, and Estrin}}]{magn-6}
\bibinfo{author}{\bibfnamefont{A.}~\bibnamefont{Vasil'ev}},
  \bibinfo{author}{\bibfnamefont{V.}~\bibnamefont{Buchel'nikov}},
  \bibinfo{author}{\bibfnamefont{T.}~\bibnamefont{Takagi}},
  \bibinfo{author}{\bibfnamefont{V.}~\bibnamefont{Khovailo}}, \bibnamefont{and}
  \bibinfo{author}{\bibfnamefont{E.}~\bibnamefont{Estrin}},
  \bibinfo{journal}{Phys. Usp.} \textbf{\bibinfo{volume}{46}},
  \bibinfo{pages}{559} (\bibinfo{year}{2003}).

\bibitem[{\citenamefont{Ma et~al.}(2008{\natexlab{b}})\citenamefont{Ma, Zhang,
  Yu, Zhu, Chen, Wu, Liu, Qu, and Li}}]{magn-7}
\bibinfo{author}{\bibfnamefont{L.}~\bibnamefont{Ma}},
  \bibinfo{author}{\bibfnamefont{H.}~\bibnamefont{Zhang}},
  \bibinfo{author}{\bibfnamefont{S.}~\bibnamefont{Yu}},
  \bibinfo{author}{\bibfnamefont{Z.}~\bibnamefont{Zhu}},
  \bibinfo{author}{\bibfnamefont{J.}~\bibnamefont{Chen}},
  \bibinfo{author}{\bibfnamefont{G.}~\bibnamefont{Wu}},
  \bibinfo{author}{\bibfnamefont{H.}~\bibnamefont{Liu}},
  \bibinfo{author}{\bibfnamefont{J.}~\bibnamefont{Qu}}, \bibnamefont{and}
  \bibinfo{author}{\bibfnamefont{Y.}~\bibnamefont{Li}}, \bibinfo{journal}{Appl.
  Phys. Lett.} \textbf{\bibinfo{volume}{92}}, \bibinfo{pages}{032509}
  (\bibinfo{year}{2008}{\natexlab{b}}).

\bibitem[{\citenamefont{Tickle and James}(1999)}]{magmech-1}
\bibinfo{author}{\bibfnamefont{R.}~\bibnamefont{Tickle}} \bibnamefont{and}
  \bibinfo{author}{\bibfnamefont{R.}~\bibnamefont{James}}, \bibinfo{journal}{J.
  Magn. Magn. Mater.} \textbf{\bibinfo{volume}{195}}, \bibinfo{pages}{627}
  (\bibinfo{year}{1999}).

\bibitem[{\citenamefont{Chmielus et~al.}(2009)\citenamefont{Chmielus, Zhang,
  Witherspoon, Dunand, and M{\"u}llner}}]{magmech-2}
\bibinfo{author}{\bibfnamefont{M.}~\bibnamefont{Chmielus}},
  \bibinfo{author}{\bibfnamefont{X.}~\bibnamefont{Zhang}},
  \bibinfo{author}{\bibfnamefont{C.}~\bibnamefont{Witherspoon}},
  \bibinfo{author}{\bibfnamefont{D.}~\bibnamefont{Dunand}}, \bibnamefont{and}
  \bibinfo{author}{\bibfnamefont{P.}~\bibnamefont{M{\"u}llner}},
  \bibinfo{journal}{Nat. Mater.} \textbf{\bibinfo{volume}{8}},
  \bibinfo{pages}{863} (\bibinfo{year}{2009}).

\bibitem[{\citenamefont{Karaman et~al.}(2006)\citenamefont{Karaman, Karaca,
  Basaran, Lagoudas, Chumlyakov, and Maier}}]{magmech-4}
\bibinfo{author}{\bibfnamefont{I.}~\bibnamefont{Karaman}},
  \bibinfo{author}{\bibfnamefont{H.}~\bibnamefont{Karaca}},
  \bibinfo{author}{\bibfnamefont{B.}~\bibnamefont{Basaran}},
  \bibinfo{author}{\bibfnamefont{D.}~\bibnamefont{Lagoudas}},
  \bibinfo{author}{\bibfnamefont{Y.}~\bibnamefont{Chumlyakov}},
  \bibnamefont{and} \bibinfo{author}{\bibfnamefont{H.}~\bibnamefont{Maier}},
  \bibinfo{journal}{Scr. Mater.} \textbf{\bibinfo{volume}{55}},
  \bibinfo{pages}{403} (\bibinfo{year}{2006}).

\bibitem[{\citenamefont{Sozinov
  et~al.}(2002{\natexlab{a}})\citenamefont{Sozinov, Likhachev, Lanska, and
  Ullakko}}]{magmech-5}
\bibinfo{author}{\bibfnamefont{A.}~\bibnamefont{Sozinov}},
  \bibinfo{author}{\bibfnamefont{A.}~\bibnamefont{Likhachev}},
  \bibinfo{author}{\bibfnamefont{N.}~\bibnamefont{Lanska}}, \bibnamefont{and}
  \bibinfo{author}{\bibfnamefont{K.}~\bibnamefont{Ullakko}},
  \bibinfo{journal}{Appl. Phys. Lett.} \textbf{\bibinfo{volume}{80}},
  \bibinfo{pages}{1746} (\bibinfo{year}{2002}{\natexlab{a}}).

\bibitem[{\citenamefont{Khan et~al.}(2007{\natexlab{a}})\citenamefont{Khan,
  Dubenko, Stadler, and Ali}}]{magcal-1}
\bibinfo{author}{\bibfnamefont{M.}~\bibnamefont{Khan}},
  \bibinfo{author}{\bibfnamefont{I.}~\bibnamefont{Dubenko}},
  \bibinfo{author}{\bibfnamefont{S.}~\bibnamefont{Stadler}}, \bibnamefont{and}
  \bibinfo{author}{\bibfnamefont{N.}~\bibnamefont{Ali}}, \bibinfo{journal}{J.
  Appl. Phys.} \textbf{\bibinfo{volume}{102}}, \bibinfo{pages}{023901}
  (\bibinfo{year}{2007}{\natexlab{a}}).

\bibitem[{\citenamefont{Khan et~al.}(2007{\natexlab{b}})\citenamefont{Khan,
  Ali, and Stadler}}]{magcal-2}
\bibinfo{author}{\bibfnamefont{M.}~\bibnamefont{Khan}},
  \bibinfo{author}{\bibfnamefont{N.}~\bibnamefont{Ali}}, \bibnamefont{and}
  \bibinfo{author}{\bibfnamefont{S.}~\bibnamefont{Stadler}},
  \bibinfo{journal}{J. Appl. Phys.} \textbf{\bibinfo{volume}{101}},
  \bibinfo{pages}{053919} (\bibinfo{year}{2007}{\natexlab{b}}).

\bibitem[{\citenamefont{Chatterjee et~al.}(2010)\citenamefont{Chatterjee, Giri,
  De, and Majumdar}}]{magcal-3}
\bibinfo{author}{\bibfnamefont{S.}~\bibnamefont{Chatterjee}},
  \bibinfo{author}{\bibfnamefont{S.}~\bibnamefont{Giri}},
  \bibinfo{author}{\bibfnamefont{S.}~\bibnamefont{De}}, \bibnamefont{and}
  \bibinfo{author}{\bibfnamefont{S.}~\bibnamefont{Majumdar}},
  \bibinfo{journal}{J. Alloys Compd.} \textbf{\bibinfo{volume}{503}},
  \bibinfo{pages}{273} (\bibinfo{year}{2010}).

\bibitem[{\citenamefont{Hu et~al.}(2001)\citenamefont{Hu, Shen, Sun, and
  Wu}}]{magcal-4}
\bibinfo{author}{\bibfnamefont{F.}~\bibnamefont{Hu}},
  \bibinfo{author}{\bibfnamefont{B.}~\bibnamefont{Shen}},
  \bibinfo{author}{\bibfnamefont{J.}~\bibnamefont{Sun}}, \bibnamefont{and}
  \bibinfo{author}{\bibfnamefont{G.}~\bibnamefont{Wu}}, \bibinfo{journal}{Phys.
  Rev. B} \textbf{\bibinfo{volume}{64}}, \bibinfo{pages}{132412}
  (\bibinfo{year}{2001}).

\bibitem[{\citenamefont{Marcos et~al.}(2002)\citenamefont{Marcos, Planes,
  Ma{\~n}osa, Casanova, Batlle, Labarta, and Mart{\'\i}nez}}]{magcal-5}
\bibinfo{author}{\bibfnamefont{J.}~\bibnamefont{Marcos}},
  \bibinfo{author}{\bibfnamefont{A.}~\bibnamefont{Planes}},
  \bibinfo{author}{\bibfnamefont{L.}~\bibnamefont{Ma{\~n}osa}},
  \bibinfo{author}{\bibfnamefont{F.}~\bibnamefont{Casanova}},
  \bibinfo{author}{\bibfnamefont{X.}~\bibnamefont{Batlle}},
  \bibinfo{author}{\bibfnamefont{A.}~\bibnamefont{Labarta}}, \bibnamefont{and}
  \bibinfo{author}{\bibfnamefont{B.}~\bibnamefont{Mart{\'\i}nez}},
  \bibinfo{journal}{Phys. Rev. B} \textbf{\bibinfo{volume}{66}},
  \bibinfo{pages}{224413} (\bibinfo{year}{2002}).

\bibitem[{\citenamefont{Cherechukin
  et~al.}(2001{\natexlab{b}})\citenamefont{Cherechukin, Dikshtein, Ermakov,
  Glebov, Koledov, Kosolapov, Shavrov, Tulaikova, Krasnoperov, and
  Takagi}}]{magmech-3}
\bibinfo{author}{\bibfnamefont{A.}~\bibnamefont{Cherechukin}},
  \bibinfo{author}{\bibfnamefont{I.}~\bibnamefont{Dikshtein}},
  \bibinfo{author}{\bibfnamefont{D.}~\bibnamefont{Ermakov}},
  \bibinfo{author}{\bibfnamefont{A.}~\bibnamefont{Glebov}},
  \bibinfo{author}{\bibfnamefont{V.}~\bibnamefont{Koledov}},
  \bibinfo{author}{\bibfnamefont{D.}~\bibnamefont{Kosolapov}},
  \bibinfo{author}{\bibfnamefont{V.}~\bibnamefont{Shavrov}},
  \bibinfo{author}{\bibfnamefont{A.}~\bibnamefont{Tulaikova}},
  \bibinfo{author}{\bibfnamefont{E.}~\bibnamefont{Krasnoperov}},
  \bibnamefont{and} \bibinfo{author}{\bibfnamefont{T.}~\bibnamefont{Takagi}},
  \bibinfo{journal}{Phys. Lett. A} \textbf{\bibinfo{volume}{291}},
  \bibinfo{pages}{175} (\bibinfo{year}{2001}{\natexlab{b}}).

\bibitem[{\citenamefont{Srivastava et~al.}(2011)\citenamefont{Srivastava, Song,
  Bhatti, and James}}]{econv}
\bibinfo{author}{\bibfnamefont{V.}~\bibnamefont{Srivastava}},
  \bibinfo{author}{\bibfnamefont{Y.}~\bibnamefont{Song}},
  \bibinfo{author}{\bibfnamefont{K.}~\bibnamefont{Bhatti}}, \bibnamefont{and}
  \bibinfo{author}{\bibfnamefont{R.}~\bibnamefont{James}},
  \bibinfo{journal}{Adv. Energy Mater.}  (\bibinfo{year}{2011}).

\bibitem[{\citenamefont{Brown et~al.}(2002)\citenamefont{Brown, Crangle,
  Kanomata, Matsumoto, Neumann, Ouladdiaf, and Ziebeck}}]{brown-modulated}
\bibinfo{author}{\bibfnamefont{P.}~\bibnamefont{Brown}},
  \bibinfo{author}{\bibfnamefont{J.}~\bibnamefont{Crangle}},
  \bibinfo{author}{\bibfnamefont{T.}~\bibnamefont{Kanomata}},
  \bibinfo{author}{\bibfnamefont{M.}~\bibnamefont{Matsumoto}},
  \bibinfo{author}{\bibfnamefont{K.}~\bibnamefont{Neumann}},
  \bibinfo{author}{\bibfnamefont{B.}~\bibnamefont{Ouladdiaf}},
  \bibnamefont{and} \bibinfo{author}{\bibfnamefont{K.}~\bibnamefont{Ziebeck}},
  \bibinfo{journal}{J. Phys.: Condens. Matter} \textbf{\bibinfo{volume}{14}},
  \bibinfo{pages}{10159} (\bibinfo{year}{2002}).

\bibitem[{\citenamefont{Kokorin et~al.}(1992)\citenamefont{Kokorin, Martynov,
  and Chernenko}}]{kokorin-modul}
\bibinfo{author}{\bibfnamefont{V.}~\bibnamefont{Kokorin}},
  \bibinfo{author}{\bibfnamefont{V.}~\bibnamefont{Martynov}}, \bibnamefont{and}
  \bibinfo{author}{\bibfnamefont{V.}~\bibnamefont{Chernenko}},
  \bibinfo{journal}{Scr. Metall. Mater.} \textbf{\bibinfo{volume}{26}},
  \bibinfo{pages}{175} (\bibinfo{year}{1992}).

\bibitem[{\citenamefont{Martynov and Kokorin}(1992)}]{martynov-modulated}
\bibinfo{author}{\bibfnamefont{V.}~\bibnamefont{Martynov}} \bibnamefont{and}
  \bibinfo{author}{\bibfnamefont{V.}~\bibnamefont{Kokorin}},
  \bibinfo{journal}{J. Phys. III} \textbf{\bibinfo{volume}{2}},
  \bibinfo{pages}{739} (\bibinfo{year}{1992}).

\bibitem[{\citenamefont{Dai et~al.}(2004)\citenamefont{Dai, Cullen, and
  Wuttig}}]{dai-eoa}
\bibinfo{author}{\bibfnamefont{L.}~\bibnamefont{Dai}},
  \bibinfo{author}{\bibfnamefont{J.}~\bibnamefont{Cullen}}, \bibnamefont{and}
  \bibinfo{author}{\bibfnamefont{M.}~\bibnamefont{Wuttig}},
  \bibinfo{journal}{J. Appl. Phys.} \textbf{\bibinfo{volume}{95}},
  \bibinfo{pages}{6957} (\bibinfo{year}{2004}).

\bibitem[{\citenamefont{Kaufmann et~al.}(2010)\citenamefont{Kaufmann,
  R{\"o}{\ss}ler, Heczko, Wuttig, Buschbeck, Schultz, and
  F{\"a}hler}}]{kaufmann-modulated}
\bibinfo{author}{\bibfnamefont{S.}~\bibnamefont{Kaufmann}},
  \bibinfo{author}{\bibfnamefont{U.}~\bibnamefont{R{\"o}{\ss}ler}},
  \bibinfo{author}{\bibfnamefont{O.}~\bibnamefont{Heczko}},
  \bibinfo{author}{\bibfnamefont{M.}~\bibnamefont{Wuttig}},
  \bibinfo{author}{\bibfnamefont{J.}~\bibnamefont{Buschbeck}},
  \bibinfo{author}{\bibfnamefont{L.}~\bibnamefont{Schultz}}, \bibnamefont{and}
  \bibinfo{author}{\bibfnamefont{S.}~\bibnamefont{F{\"a}hler}},
  \bibinfo{journal}{Phys. Rev. Lett.} \textbf{\bibinfo{volume}{104}},
  \bibinfo{pages}{145702} (\bibinfo{year}{2010}).

\bibitem[{\citenamefont{Wedel et~al.}(1999)\citenamefont{Wedel, Suzuki,
  Murakami, Wedel, Suzuki, Shindo, and Itagaki}}]{wedel-modulated}
\bibinfo{author}{\bibfnamefont{B.}~\bibnamefont{Wedel}},
  \bibinfo{author}{\bibfnamefont{M.}~\bibnamefont{Suzuki}},
  \bibinfo{author}{\bibfnamefont{Y.}~\bibnamefont{Murakami}},
  \bibinfo{author}{\bibfnamefont{C.}~\bibnamefont{Wedel}},
  \bibinfo{author}{\bibfnamefont{T.}~\bibnamefont{Suzuki}},
  \bibinfo{author}{\bibfnamefont{D.}~\bibnamefont{Shindo}}, \bibnamefont{and}
  \bibinfo{author}{\bibfnamefont{K.}~\bibnamefont{Itagaki}},
  \bibinfo{journal}{J. Alloys Compd.} \textbf{\bibinfo{volume}{290}},
  \bibinfo{pages}{137} (\bibinfo{year}{1999}).

\bibitem[{\citenamefont{Sozinov
  et~al.}(2002{\natexlab{b}})\citenamefont{Sozinov, Likhachev, and
  Ullakko}}]{sozinov-aniso}
\bibinfo{author}{\bibfnamefont{A.}~\bibnamefont{Sozinov}},
  \bibinfo{author}{\bibfnamefont{A.}~\bibnamefont{Likhachev}},
  \bibnamefont{and} \bibinfo{author}{\bibfnamefont{K.}~\bibnamefont{Ullakko}},
  \bibinfo{journal}{IEEE Trans. Magn.} \textbf{\bibinfo{volume}{38}},
  \bibinfo{pages}{2814} (\bibinfo{year}{2002}{\natexlab{b}}).

\bibitem[{\citenamefont{Ayuela et~al.}(2002)\citenamefont{Ayuela, Enkovaara,
  and Nieminen}}]{enkovaara-coa}
\bibinfo{author}{\bibfnamefont{A.}~\bibnamefont{Ayuela}},
  \bibinfo{author}{\bibfnamefont{J.}~\bibnamefont{Enkovaara}},
  \bibnamefont{and} \bibinfo{author}{\bibfnamefont{R.}~\bibnamefont{Nieminen}},
  \bibinfo{journal}{J. Phys.: Condens. Matter} \textbf{\bibinfo{volume}{14}},
  \bibinfo{pages}{5325} (\bibinfo{year}{2002}).

\bibitem[{\citenamefont{Entel et~al.}(2006)\citenamefont{Entel, Buchelnikov,
  Khovailo, Zayak, Adeagbo, Gruner, Herper, and Wassermann}}]{entel-review}
\bibinfo{author}{\bibfnamefont{P.}~\bibnamefont{Entel}},
  \bibinfo{author}{\bibfnamefont{V.}~\bibnamefont{Buchelnikov}},
  \bibinfo{author}{\bibfnamefont{V.}~\bibnamefont{Khovailo}},
  \bibinfo{author}{\bibfnamefont{A.}~\bibnamefont{Zayak}},
  \bibinfo{author}{\bibfnamefont{W.}~\bibnamefont{Adeagbo}},
  \bibinfo{author}{\bibfnamefont{M.}~\bibnamefont{Gruner}},
  \bibinfo{author}{\bibfnamefont{H.}~\bibnamefont{Herper}}, \bibnamefont{and}
  \bibinfo{author}{\bibfnamefont{E.}~\bibnamefont{Wassermann}},
  \bibinfo{journal}{J. Phys. D: Appl. Phys.} \textbf{\bibinfo{volume}{39}},
  \bibinfo{pages}{865} (\bibinfo{year}{2006}).

\bibitem[{\citenamefont{Godlevsky and Rabe}(2001)}]{rabe-coa}
\bibinfo{author}{\bibfnamefont{V.}~\bibnamefont{Godlevsky}} \bibnamefont{and}
  \bibinfo{author}{\bibfnamefont{K.}~\bibnamefont{Rabe}},
  \bibinfo{journal}{Phys. Rev. B} \textbf{\bibinfo{volume}{63}},
  \bibinfo{pages}{134407} (\bibinfo{year}{2001}).

\bibitem[{\citenamefont{Zayak et~al.}(2003)\citenamefont{Zayak, Entel,
  Enkovaara, Ayuela, and Nieminen}}]{ayuela-coa}
\bibinfo{author}{\bibfnamefont{A.}~\bibnamefont{Zayak}},
  \bibinfo{author}{\bibfnamefont{P.}~\bibnamefont{Entel}},
  \bibinfo{author}{\bibfnamefont{J.}~\bibnamefont{Enkovaara}},
  \bibinfo{author}{\bibfnamefont{A.}~\bibnamefont{Ayuela}}, \bibnamefont{and}
  \bibinfo{author}{\bibfnamefont{R.}~\bibnamefont{Nieminen}},
  \bibinfo{journal}{J. Phys.: Condens. Matter} \textbf{\bibinfo{volume}{15}},
  \bibinfo{pages}{159} (\bibinfo{year}{2003}).

\bibitem[{\citenamefont{Ayuela et~al.}(1999)\citenamefont{Ayuela, Enkovaara,
  Ullakko, and Nieminen}}]{ayuela-coa2}
\bibinfo{author}{\bibfnamefont{A.}~\bibnamefont{Ayuela}},
  \bibinfo{author}{\bibfnamefont{J.}~\bibnamefont{Enkovaara}},
  \bibinfo{author}{\bibfnamefont{K.}~\bibnamefont{Ullakko}}, \bibnamefont{and}
  \bibinfo{author}{\bibfnamefont{R.}~\bibnamefont{Nieminen}},
  \bibinfo{journal}{J. Phys.: Condens. Matter} \textbf{\bibinfo{volume}{11}},
  \bibinfo{pages}{2017} (\bibinfo{year}{1999}).

\bibitem[{\citenamefont{{\"O}zdemir~Kart
  et~al.}(2008)\citenamefont{{\"O}zdemir~Kart, Uludogan, Karaman, and
  T.}}]{kart-coa}
\bibinfo{author}{\bibfnamefont{S.}~\bibnamefont{{\"O}zdemir~Kart}},
  \bibinfo{author}{\bibfnamefont{M.}~\bibnamefont{Uludogan}},
  \bibinfo{author}{\bibfnamefont{I.}~\bibnamefont{Karaman}}, \bibnamefont{and}
  \bibinfo{author}{\bibfnamefont{C.}~\bibnamefont{T.}}, \bibinfo{journal}{Phys.
  Status Solidi A} \textbf{\bibinfo{volume}{205}}, \bibinfo{pages}{1026}
  (\bibinfo{year}{2008}).

\bibitem[{\citenamefont{Enkovaara et~al.}(2003)\citenamefont{Enkovaara, Heczko,
  Ayuela, and Nieminen}}]{enkovaara-coa2}
\bibinfo{author}{\bibfnamefont{J.}~\bibnamefont{Enkovaara}},
  \bibinfo{author}{\bibfnamefont{O.}~\bibnamefont{Heczko}},
  \bibinfo{author}{\bibfnamefont{A.}~\bibnamefont{Ayuela}}, \bibnamefont{and}
  \bibinfo{author}{\bibfnamefont{R.}~\bibnamefont{Nieminen}},
  \bibinfo{journal}{Phys. Rev. B} \textbf{\bibinfo{volume}{67}},
  \bibinfo{pages}{212405} (\bibinfo{year}{2003}).

\bibitem[{\citenamefont{Hu et~al.}(2010)\citenamefont{Hu, Li, Kulkova, Yang,
  Johansson, and Vitos}}]{hu-x}
\bibinfo{author}{\bibfnamefont{Q.}~\bibnamefont{Hu}},
  \bibinfo{author}{\bibfnamefont{C.}~\bibnamefont{Li}},
  \bibinfo{author}{\bibfnamefont{S.}~\bibnamefont{Kulkova}},
  \bibinfo{author}{\bibfnamefont{R.}~\bibnamefont{Yang}},
  \bibinfo{author}{\bibfnamefont{B.}~\bibnamefont{Johansson}},
  \bibnamefont{and} \bibinfo{author}{\bibfnamefont{L.}~\bibnamefont{Vitos}},
  \bibinfo{journal}{Phys. Rev. B} \textbf{\bibinfo{volume}{81}},
  \bibinfo{pages}{064108} (\bibinfo{year}{2010}).

\bibitem[{\citenamefont{Ye et~al.}(2010)\citenamefont{Ye, Kimura, Miura,
  Shirai, Cui, Shimada, Namatame, Taniguchi, Ueda, Kobayashi et~al.}}]{Ni2MnSn}
\bibinfo{author}{\bibfnamefont{M.}~\bibnamefont{Ye}},
  \bibinfo{author}{\bibfnamefont{A.}~\bibnamefont{Kimura}},
  \bibinfo{author}{\bibfnamefont{Y.}~\bibnamefont{Miura}},
  \bibinfo{author}{\bibfnamefont{M.}~\bibnamefont{Shirai}},
  \bibinfo{author}{\bibfnamefont{Y.}~\bibnamefont{Cui}},
  \bibinfo{author}{\bibfnamefont{K.}~\bibnamefont{Shimada}},
  \bibinfo{author}{\bibfnamefont{H.}~\bibnamefont{Namatame}},
  \bibinfo{author}{\bibfnamefont{M.}~\bibnamefont{Taniguchi}},
  \bibinfo{author}{\bibfnamefont{S.}~\bibnamefont{Ueda}},
  \bibinfo{author}{\bibfnamefont{K.}~\bibnamefont{Kobayashi}},
  \bibnamefont{et~al.}, \bibinfo{journal}{Phy. Rev. Lett.}
  \textbf{\bibinfo{volume}{104}}, \bibinfo{pages}{176401}
  (\bibinfo{year}{2010}).

\bibitem[{\citenamefont{Zayak et~al.}(2005)\citenamefont{Zayak, Entel, Rabe,
  Adeagbo, and Acet}}]{rabe-vib}
\bibinfo{author}{\bibfnamefont{A.}~\bibnamefont{Zayak}},
  \bibinfo{author}{\bibfnamefont{P.}~\bibnamefont{Entel}},
  \bibinfo{author}{\bibfnamefont{K.}~\bibnamefont{Rabe}},
  \bibinfo{author}{\bibfnamefont{W.}~\bibnamefont{Adeagbo}}, \bibnamefont{and}
  \bibinfo{author}{\bibfnamefont{M.}~\bibnamefont{Acet}},
  \bibinfo{journal}{Phys. Rev. B} \textbf{\bibinfo{volume}{72}},
  \bibinfo{pages}{054113} (\bibinfo{year}{2005}).

\bibitem[{\citenamefont{Bungaro et~al.}(2003)\citenamefont{Bungaro, Rabe, and
  Dal~Corso}}]{rabe-phon}
\bibinfo{author}{\bibfnamefont{C.}~\bibnamefont{Bungaro}},
  \bibinfo{author}{\bibfnamefont{K.}~\bibnamefont{Rabe}}, \bibnamefont{and}
  \bibinfo{author}{\bibfnamefont{A.}~\bibnamefont{Dal~Corso}},
  \bibinfo{journal}{Phys. Rev. B} \textbf{\bibinfo{volume}{68}},
  \bibinfo{pages}{134104} (\bibinfo{year}{2003}).

\bibitem[{\citenamefont{Ming et~al.}(2009)\citenamefont{Ming, Liu, Zhang, Zhao,
  and Yao}}]{el-ph}
\bibinfo{author}{\bibfnamefont{W.}~\bibnamefont{Ming}},
  \bibinfo{author}{\bibfnamefont{Y.}~\bibnamefont{Liu}},
  \bibinfo{author}{\bibfnamefont{W.}~\bibnamefont{Zhang}},
  \bibinfo{author}{\bibfnamefont{J.}~\bibnamefont{Zhao}}, \bibnamefont{and}
  \bibinfo{author}{\bibfnamefont{Y.}~\bibnamefont{Yao}}, \bibinfo{journal}{J.
  Phys.: Condens. Matter} \textbf{\bibinfo{volume}{21}},
  \bibinfo{pages}{075501} (\bibinfo{year}{2009}).

\bibitem[{\citenamefont{Uijttewaal et~al.}(2009)\citenamefont{Uijttewaal,
  Hickel, Neugebauer, Gruner, and Entel}}]{phase}
\bibinfo{author}{\bibfnamefont{M.}~\bibnamefont{Uijttewaal}},
  \bibinfo{author}{\bibfnamefont{T.}~\bibnamefont{Hickel}},
  \bibinfo{author}{\bibfnamefont{J.}~\bibnamefont{Neugebauer}},
  \bibinfo{author}{\bibfnamefont{M.}~\bibnamefont{Gruner}}, \bibnamefont{and}
  \bibinfo{author}{\bibfnamefont{P.}~\bibnamefont{Entel}},
  \bibinfo{journal}{Physical review letters} \textbf{\bibinfo{volume}{102}},
  \bibinfo{pages}{35702} (\bibinfo{year}{2009}).

\bibitem[{\citenamefont{{\c{S}}a{\c{s}}{\i}o{\u{g}}lu
  et~al.}(2004)\citenamefont{{\c{S}}a{\c{s}}{\i}o{\u{g}}lu, Sandratskii, and
  Bruno}}]{heusler-rkky2}
\bibinfo{author}{\bibfnamefont{E.}~\bibnamefont{{\c{S}}a{\c{s}}{\i}o{\u{g}}lu}%
}, \bibinfo{author}{\bibfnamefont{L.}~\bibnamefont{Sandratskii}},
  \bibnamefont{and} \bibinfo{author}{\bibfnamefont{P.}~\bibnamefont{Bruno}},
  \bibinfo{journal}{Phys. Rev. B} \textbf{\bibinfo{volume}{70}},
  \bibinfo{pages}{024427} (\bibinfo{year}{2004}).

\bibitem[{\citenamefont{Rusz et~al.}(2006)\citenamefont{Rusz, Bergqvist,
  Kudrnovsk{\`y}, and Turek}}]{heusler-rkky4}
\bibinfo{author}{\bibfnamefont{J.}~\bibnamefont{Rusz}},
  \bibinfo{author}{\bibfnamefont{L.}~\bibnamefont{Bergqvist}},
  \bibinfo{author}{\bibfnamefont{J.}~\bibnamefont{Kudrnovsk{\`y}}},
  \bibnamefont{and} \bibinfo{author}{\bibfnamefont{I.}~\bibnamefont{Turek}},
  \bibinfo{journal}{Phys. Rev. B} \textbf{\bibinfo{volume}{73}},
  \bibinfo{pages}{214412} (\bibinfo{year}{2006}).

\bibitem[{\citenamefont{{\c{S}}a{\c{s}}{\i}o{\u{g}}lu
  et~al.}(2005)\citenamefont{{\c{S}}a{\c{s}}{\i}o{\u{g}}lu, Sandratskii, and
  Bruno}}]{Tc}
\bibinfo{author}{\bibfnamefont{E.}~\bibnamefont{{\c{S}}a{\c{s}}{\i}o{\u{g}}lu}%
}, \bibinfo{author}{\bibfnamefont{L.}~\bibnamefont{Sandratskii}},
  \bibnamefont{and} \bibinfo{author}{\bibfnamefont{P.}~\bibnamefont{Bruno}},
  \bibinfo{journal}{Phys. Rev. B} \textbf{\bibinfo{volume}{71}},
  \bibinfo{pages}{214412} (\bibinfo{year}{2005}).

\bibitem[{\citenamefont{Brown et~al.}(1999)\citenamefont{Brown, Bargawi,
  Crangle, Neumann, and Ziebeck}}]{brown-jahnteller}
\bibinfo{author}{\bibfnamefont{P.}~\bibnamefont{Brown}},
  \bibinfo{author}{\bibfnamefont{A.}~\bibnamefont{Bargawi}},
  \bibinfo{author}{\bibfnamefont{J.}~\bibnamefont{Crangle}},
  \bibinfo{author}{\bibfnamefont{K.}~\bibnamefont{Neumann}}, \bibnamefont{and}
  \bibinfo{author}{\bibfnamefont{K.}~\bibnamefont{Ziebeck}},
  \bibinfo{journal}{J. Phys.: Condens. Matter} \textbf{\bibinfo{volume}{11}},
  \bibinfo{pages}{4715} (\bibinfo{year}{1999}).

\bibitem[{\citenamefont{Fujii et~al.}(1989)\citenamefont{Fujii, Ishida, and
  Asano}}]{fujii-jahnteller}
\bibinfo{author}{\bibfnamefont{S.}~\bibnamefont{Fujii}},
  \bibinfo{author}{\bibfnamefont{S.}~\bibnamefont{Ishida}}, \bibnamefont{and}
  \bibinfo{author}{\bibfnamefont{S.}~\bibnamefont{Asano}}, \bibinfo{journal}{J.
  Phys. Soc. Jpn.} \textbf{\bibinfo{volume}{58}}, \bibinfo{pages}{3657}
  (\bibinfo{year}{1989}).

\bibitem[{\citenamefont{Opeil et~al.}(2008)\citenamefont{Opeil, Mihaila,
  Schulze, Ma{\~n}osa, Planes, Hults, Fisher, Riseborough, Littlewood, Smith
  et~al.}}]{opeil-pre}
\bibinfo{author}{\bibfnamefont{C.}~\bibnamefont{Opeil}},
  \bibinfo{author}{\bibfnamefont{B.}~\bibnamefont{Mihaila}},
  \bibinfo{author}{\bibfnamefont{R.}~\bibnamefont{Schulze}},
  \bibinfo{author}{\bibfnamefont{L.}~\bibnamefont{Ma{\~n}osa}},
  \bibinfo{author}{\bibfnamefont{A.}~\bibnamefont{Planes}},
  \bibinfo{author}{\bibfnamefont{W.}~\bibnamefont{Hults}},
  \bibinfo{author}{\bibfnamefont{R.}~\bibnamefont{Fisher}},
  \bibinfo{author}{\bibfnamefont{P.}~\bibnamefont{Riseborough}},
  \bibinfo{author}{\bibfnamefont{P.}~\bibnamefont{Littlewood}},
  \bibinfo{author}{\bibfnamefont{J.}~\bibnamefont{Smith}},
  \bibnamefont{et~al.}, \bibinfo{journal}{Phys. Rev. Lett.}
  \textbf{\bibinfo{volume}{100}}, \bibinfo{pages}{165703}
  (\bibinfo{year}{2008}).

\bibitem[{\citenamefont{Zheludev et~al.}(1996)\citenamefont{Zheludev, Shapiro,
  Wochner, and Tanner}}]{zheludev-fs}
\bibinfo{author}{\bibfnamefont{A.}~\bibnamefont{Zheludev}},
  \bibinfo{author}{\bibfnamefont{S.}~\bibnamefont{Shapiro}},
  \bibinfo{author}{\bibfnamefont{P.}~\bibnamefont{Wochner}}, \bibnamefont{and}
  \bibinfo{author}{\bibfnamefont{L.}~\bibnamefont{Tanner}},
  \bibinfo{journal}{Phys. Rev. B} \textbf{\bibinfo{volume}{54}},
  \bibinfo{pages}{15045} (\bibinfo{year}{1996}).

\bibitem[{\citenamefont{Zayak et~al.}(2006)\citenamefont{Zayak, Adeagbo, Entel,
  and Rabe}}]{rabe-eoa}
\bibinfo{author}{\bibfnamefont{A.}~\bibnamefont{Zayak}},
  \bibinfo{author}{\bibfnamefont{W.}~\bibnamefont{Adeagbo}},
  \bibinfo{author}{\bibfnamefont{P.}~\bibnamefont{Entel}}, \bibnamefont{and}
  \bibinfo{author}{\bibfnamefont{K.}~\bibnamefont{Rabe}},
  \bibinfo{journal}{Appl. Phys. Lett.} \textbf{\bibinfo{volume}{88}},
  \bibinfo{pages}{111903} (\bibinfo{year}{2006}).

\bibitem[{\citenamefont{Hu et~al.}(2009)\citenamefont{Hu, Li, Yang, Kulkova,
  Bazhanov, Johansson, and Vitos}}]{hu-eoa}
\bibinfo{author}{\bibfnamefont{Q.}~\bibnamefont{Hu}},
  \bibinfo{author}{\bibfnamefont{C.}~\bibnamefont{Li}},
  \bibinfo{author}{\bibfnamefont{R.}~\bibnamefont{Yang}},
  \bibinfo{author}{\bibfnamefont{S.}~\bibnamefont{Kulkova}},
  \bibinfo{author}{\bibfnamefont{D.}~\bibnamefont{Bazhanov}},
  \bibinfo{author}{\bibfnamefont{B.}~\bibnamefont{Johansson}},
  \bibnamefont{and} \bibinfo{author}{\bibfnamefont{L.}~\bibnamefont{Vitos}},
  \bibinfo{journal}{Phys. Rev. B} \textbf{\bibinfo{volume}{79}},
  \bibinfo{pages}{144112} (\bibinfo{year}{2009}).

\bibitem[{\citenamefont{Lanska et~al.}(2004)\citenamefont{Lanska, Soderberg,
  Sozinov, Ge, Ullakko, and Lindroos}}]{lanska-coa}
\bibinfo{author}{\bibfnamefont{N.}~\bibnamefont{Lanska}},
  \bibinfo{author}{\bibfnamefont{O.}~\bibnamefont{Soderberg}},
  \bibinfo{author}{\bibfnamefont{A.}~\bibnamefont{Sozinov}},
  \bibinfo{author}{\bibfnamefont{Y.}~\bibnamefont{Ge}},
  \bibinfo{author}{\bibfnamefont{K.}~\bibnamefont{Ullakko}}, \bibnamefont{and}
  \bibinfo{author}{\bibfnamefont{V.}~\bibnamefont{Lindroos}},
  \bibinfo{journal}{J. Appl. Phys.} \textbf{\bibinfo{volume}{95}},
  \bibinfo{pages}{8074} (\bibinfo{year}{2004}).

\bibitem[{\citenamefont{Cococcioni and De~Gironcoli}(2005)}]{Ucalc}
\bibinfo{author}{\bibfnamefont{M.}~\bibnamefont{Cococcioni}} \bibnamefont{and}
  \bibinfo{author}{\bibfnamefont{S.}~\bibnamefont{De~Gironcoli}},
  \bibinfo{journal}{Phys. Rev. B} \textbf{\bibinfo{volume}{71}},
  \bibinfo{pages}{35105} (\bibinfo{year}{2005}).

\bibitem[{\citenamefont{Anisimov et~al.}(1991)\citenamefont{Anisimov, Zaanen,
  and Andersen}}]{anisimov-1991}
\bibinfo{author}{\bibfnamefont{V.}~\bibnamefont{Anisimov}},
  \bibinfo{author}{\bibfnamefont{J.}~\bibnamefont{Zaanen}}, \bibnamefont{and}
  \bibinfo{author}{\bibfnamefont{O.}~\bibnamefont{Andersen}},
  \bibinfo{journal}{Phys. Rev. B} \textbf{\bibinfo{volume}{44}},
  \bibinfo{pages}{943} (\bibinfo{year}{1991}).

\bibitem[{\citenamefont{Anisimov et~al.}(1993)\citenamefont{Anisimov, Solovyev,
  Korotin, Czy{\.z}yk, and Sawatzky}}]{anisimov-1993}
\bibinfo{author}{\bibfnamefont{V.}~\bibnamefont{Anisimov}},
  \bibinfo{author}{\bibfnamefont{I.}~\bibnamefont{Solovyev}},
  \bibinfo{author}{\bibfnamefont{M.}~\bibnamefont{Korotin}},
  \bibinfo{author}{\bibfnamefont{M.}~\bibnamefont{Czy{\.z}yk}},
  \bibnamefont{and} \bibinfo{author}{\bibfnamefont{G.}~\bibnamefont{Sawatzky}},
  \bibinfo{journal}{Phys. Rev. B} \textbf{\bibinfo{volume}{48}},
  \bibinfo{pages}{16929} (\bibinfo{year}{1993}).

\bibitem[{\citenamefont{Mazin and Anisimov}(1997)}]{mazin-1997}
\bibinfo{author}{\bibfnamefont{I.}~\bibnamefont{Mazin}} \bibnamefont{and}
  \bibinfo{author}{\bibfnamefont{V.}~\bibnamefont{Anisimov}},
  \bibinfo{journal}{Phys. Rev. B} \textbf{\bibinfo{volume}{55}},
  \bibinfo{pages}{12822} (\bibinfo{year}{1997}).

\bibitem[{\citenamefont{Solovyev et~al.}(1998)\citenamefont{Solovyev,
  Liechtenstein, and Terakura}}]{solovyev-1998}
\bibinfo{author}{\bibfnamefont{I.}~\bibnamefont{Solovyev}},
  \bibinfo{author}{\bibfnamefont{A.}~\bibnamefont{Liechtenstein}},
  \bibnamefont{and} \bibinfo{author}{\bibfnamefont{K.}~\bibnamefont{Terakura}},
  \bibinfo{journal}{J. Magn. Magn. Mater.} \textbf{\bibinfo{volume}{185}},
  \bibinfo{pages}{118} (\bibinfo{year}{1998}).

\bibitem[{\citenamefont{Anderson}(1961)}]{anderson-imp}
\bibinfo{author}{\bibfnamefont{P.}~\bibnamefont{Anderson}},
  \bibinfo{journal}{Phys. Rev.} \textbf{\bibinfo{volume}{124}},
  \bibinfo{pages}{41} (\bibinfo{year}{1961}).

\bibitem[{\citenamefont{Giannozzi et~al.}(2009)\citenamefont{Giannozzi, Baroni,
  Bonini, Calandra, Car, Cavazzoni, Ceresoli, Chiarotti, Cococcioni, Dabo
  et~al.}}]{espresso}
\bibinfo{author}{\bibfnamefont{P.}~\bibnamefont{Giannozzi}},
  \bibinfo{author}{\bibfnamefont{S.}~\bibnamefont{Baroni}},
  \bibinfo{author}{\bibfnamefont{N.}~\bibnamefont{Bonini}},
  \bibinfo{author}{\bibfnamefont{M.}~\bibnamefont{Calandra}},
  \bibinfo{author}{\bibfnamefont{R.}~\bibnamefont{Car}},
  \bibinfo{author}{\bibfnamefont{C.}~\bibnamefont{Cavazzoni}},
  \bibinfo{author}{\bibfnamefont{D.}~\bibnamefont{Ceresoli}},
  \bibinfo{author}{\bibfnamefont{G.}~\bibnamefont{Chiarotti}},
  \bibinfo{author}{\bibfnamefont{M.}~\bibnamefont{Cococcioni}},
  \bibinfo{author}{\bibfnamefont{I.}~\bibnamefont{Dabo}}, \bibnamefont{et~al.},
  \bibinfo{journal}{J. Phys. Condens. Matter} \textbf{\bibinfo{volume}{21}},
  \bibinfo{pages}{395502} (\bibinfo{year}{2009}).

\bibitem[{\citenamefont{Perdew et~al.}(1996)\citenamefont{Perdew, Burke, and
  Ernzerhof}}]{pbe}
\bibinfo{author}{\bibfnamefont{J.}~\bibnamefont{Perdew}},
  \bibinfo{author}{\bibfnamefont{K.}~\bibnamefont{Burke}}, \bibnamefont{and}
  \bibinfo{author}{\bibfnamefont{M.}~\bibnamefont{Ernzerhof}},
  \bibinfo{journal}{Phys. Rev. Lett} \textbf{\bibinfo{volume}{77}},
  \bibinfo{pages}{3865} (\bibinfo{year}{1996}).

\bibitem[{\citenamefont{Vanderbilt}(1990)}]{vanderbilt}
\bibinfo{author}{\bibfnamefont{D.}~\bibnamefont{Vanderbilt}},
  \bibinfo{journal}{Phys. Rev. B} \textbf{\bibinfo{volume}{41}},
  \bibinfo{pages}{7892} (\bibinfo{year}{1990}).

\bibitem[{\citenamefont{Monkhorst and Pack}(1976)}]{BZ}
\bibinfo{author}{\bibfnamefont{H.}~\bibnamefont{Monkhorst}} \bibnamefont{and}
  \bibinfo{author}{\bibfnamefont{J.}~\bibnamefont{Pack}},
  \bibinfo{journal}{Phys. Rev. B} \textbf{\bibinfo{volume}{13}},
  \bibinfo{pages}{5188} (\bibinfo{year}{1976}).

\bibitem[{\citenamefont{Methfessel and Paxton}(1989)}]{mp}
\bibinfo{author}{\bibfnamefont{M.}~\bibnamefont{Methfessel}} \bibnamefont{and}
  \bibinfo{author}{\bibfnamefont{A.}~\bibnamefont{Paxton}},
  \bibinfo{journal}{Phys. Rev. B} \textbf{\bibinfo{volume}{40}},
  \bibinfo{pages}{3616} (\bibinfo{year}{1989}).

\bibitem[{\citenamefont{Himmetoglu et~al.}(2011)\citenamefont{Himmetoglu,
  Wentzcovitch, and Cococcioni}}]{hwc}
\bibinfo{author}{\bibfnamefont{B.}~\bibnamefont{Himmetoglu}},
  \bibinfo{author}{\bibfnamefont{R.~M.} \bibnamefont{Wentzcovitch}},
  \bibnamefont{and}
  \bibinfo{author}{\bibfnamefont{M.}~\bibnamefont{Cococcioni}},
  \bibinfo{journal}{Phys. Rev. B} \textbf{\bibinfo{volume}{84}},
  \bibinfo{pages}{115108} (\bibinfo{year}{2011}).

\bibitem[{\citenamefont{Campo~Jr and Cococcioni}(2010)}]{HubV}
\bibinfo{author}{\bibfnamefont{V.}~\bibnamefont{Campo~Jr}} \bibnamefont{and}
  \bibinfo{author}{\bibfnamefont{M.}~\bibnamefont{Cococcioni}},
  \bibinfo{journal}{J. Phys. Condens. Matter} \textbf{\bibinfo{volume}{22}},
  \bibinfo{pages}{055602} (\bibinfo{year}{2010}).

\bibitem[{\citenamefont{Kokalj}(2003)}]{xcrysden}
\bibinfo{author}{\bibfnamefont{A.}~\bibnamefont{Kokalj}},
  \bibinfo{journal}{Comp. Mater. Sci.} \textbf{\bibinfo{volume}{28}},
  \bibinfo{pages}{155} (\bibinfo{year}{2003}).

\bibitem[{\citenamefont{Webster}(1969)}]{heusler}
\bibinfo{author}{\bibfnamefont{P.}~\bibnamefont{Webster}},
  \bibinfo{journal}{Contemp. Phys.} \textbf{\bibinfo{volume}{10}},
  \bibinfo{pages}{559} (\bibinfo{year}{1969}).

\bibitem[{\citenamefont{Kulik et~al.}(2006)\citenamefont{Kulik, Cococcioni,
  Scherlis, and Marzari}}]{self-cons-u}
\bibinfo{author}{\bibfnamefont{H.}~\bibnamefont{Kulik}},
  \bibinfo{author}{\bibfnamefont{M.}~\bibnamefont{Cococcioni}},
  \bibinfo{author}{\bibfnamefont{D.}~\bibnamefont{Scherlis}}, \bibnamefont{and}
  \bibinfo{author}{\bibfnamefont{N.}~\bibnamefont{Marzari}},
  \bibinfo{journal}{Phys. Rev. Lett.} \textbf{\bibinfo{volume}{97}},
  \bibinfo{pages}{103001} (\bibinfo{year}{2006}).

\bibitem[{\citenamefont{Hsu et~al.}(2009)\citenamefont{Hsu, Umemoto,
  Cococcioni, and Wentzcovitch}}]{str-cons-u}
\bibinfo{author}{\bibfnamefont{H.}~\bibnamefont{Hsu}},
  \bibinfo{author}{\bibfnamefont{K.}~\bibnamefont{Umemoto}},
  \bibinfo{author}{\bibfnamefont{M.}~\bibnamefont{Cococcioni}},
  \bibnamefont{and}
  \bibinfo{author}{\bibfnamefont{R.}~\bibnamefont{Wentzcovitch}},
  \bibinfo{journal}{Phys. Rev. B} \textbf{\bibinfo{volume}{79}},
  \bibinfo{pages}{125124} (\bibinfo{year}{2009}).

\bibitem[{\citenamefont{Klaer et~al.}(2011)\citenamefont{Klaer, Eichhorn,
  Jakob, and Elmers}}]{xray}
\bibinfo{author}{\bibfnamefont{P.}~\bibnamefont{Klaer}},
  \bibinfo{author}{\bibfnamefont{T.}~\bibnamefont{Eichhorn}},
  \bibinfo{author}{\bibfnamefont{G.}~\bibnamefont{Jakob}}, \bibnamefont{and}
  \bibinfo{author}{\bibfnamefont{H.}~\bibnamefont{Elmers}},
  \bibinfo{journal}{Phys. Rev. B} \textbf{\bibinfo{volume}{83}},
  \bibinfo{pages}{214419} (\bibinfo{year}{2011}).

\bibitem[{\citenamefont{Anderson}(1959)}]{anderson-tJ}
\bibinfo{author}{\bibfnamefont{P.}~\bibnamefont{Anderson}},
  \bibinfo{journal}{Phys. Rev.} \textbf{\bibinfo{volume}{115}},
  \bibinfo{pages}{2} (\bibinfo{year}{1959}).

\bibitem[{\citenamefont{Spa{\l}ek}(1988)}]{spalek-tJ}
\bibinfo{author}{\bibfnamefont{J.}~\bibnamefont{Spa{\l}ek}},
  \bibinfo{journal}{Phys. Rev. B} \textbf{\bibinfo{volume}{37}},
  \bibinfo{pages}{533} (\bibinfo{year}{1988}).

\bibitem[{\citenamefont{Ruderman and Kittel}(1954)}]{rkky}
\bibinfo{author}{\bibfnamefont{M.}~\bibnamefont{Ruderman}} \bibnamefont{and}
  \bibinfo{author}{\bibfnamefont{C.}~\bibnamefont{Kittel}},
  \bibinfo{journal}{Phys. Rev.} \textbf{\bibinfo{volume}{96}},
  \bibinfo{pages}{99} (\bibinfo{year}{1954}).

\bibitem[{\citenamefont{Anderson}(1950)}]{anderson-sxc}
\bibinfo{author}{\bibfnamefont{P.}~\bibnamefont{Anderson}},
  \bibinfo{journal}{Phys. Rev.} \textbf{\bibinfo{volume}{79}},
  \bibinfo{pages}{350} (\bibinfo{year}{1950}).

\bibitem[{\citenamefont{Khoi et~al.}(1978)\citenamefont{Khoi, Veillet, and
  Campbell}}]{khoi-rkky}
\bibinfo{author}{\bibfnamefont{L.}~\bibnamefont{Khoi}},
  \bibinfo{author}{\bibfnamefont{P.}~\bibnamefont{Veillet}}, \bibnamefont{and}
  \bibinfo{author}{\bibfnamefont{I.}~\bibnamefont{Campbell}},
  \bibinfo{journal}{J. Phys. F: Met. Phys.} \textbf{\bibinfo{volume}{8}},
  \bibinfo{pages}{1811} (\bibinfo{year}{1978}).

\bibitem[{\citenamefont{{\c{S}}a{\c{s}}{\i}o{\u{g}}lu
  et~al.}(2008)\citenamefont{{\c{S}}a{\c{s}}{\i}o{\u{g}}lu, Sandratskii, and
  Bruno}}]{heusler-rkky1}
\bibinfo{author}{\bibfnamefont{E.}~\bibnamefont{{\c{S}}a{\c{s}}{\i}o{\u{g}}lu}%
}, \bibinfo{author}{\bibfnamefont{L.}~\bibnamefont{Sandratskii}},
  \bibnamefont{and} \bibinfo{author}{\bibfnamefont{P.}~\bibnamefont{Bruno}},
  \bibinfo{journal}{Phys. Rev. B} \textbf{\bibinfo{volume}{77}},
  \bibinfo{pages}{064417} (\bibinfo{year}{2008}).

\bibitem[{\citenamefont{Kurtulus et~al.}(2005)\citenamefont{Kurtulus,
  Dronskowski, Samolyuk, and Antropov}}]{heusler-rkky3}
\bibinfo{author}{\bibfnamefont{Y.}~\bibnamefont{Kurtulus}},
  \bibinfo{author}{\bibfnamefont{R.}~\bibnamefont{Dronskowski}},
  \bibinfo{author}{\bibfnamefont{G.}~\bibnamefont{Samolyuk}}, \bibnamefont{and}
  \bibinfo{author}{\bibfnamefont{V.}~\bibnamefont{Antropov}},
  \bibinfo{journal}{Phys. Rev. B} \textbf{\bibinfo{volume}{71}},
  \bibinfo{pages}{014425} (\bibinfo{year}{2005}).

\bibitem[{\citenamefont{Alexander and Anderson}(1964)}]{anderson-imp2}
\bibinfo{author}{\bibfnamefont{S.}~\bibnamefont{Alexander}} \bibnamefont{and}
  \bibinfo{author}{\bibfnamefont{P.}~\bibnamefont{Anderson}},
  \bibinfo{journal}{Phys. Rev.} \textbf{\bibinfo{volume}{133}},
  \bibinfo{pages}{A1594} (\bibinfo{year}{1964}).

\bibitem[{\citenamefont{Schrieffer and Wolff}(1966)}]{SW}
\bibinfo{author}{\bibfnamefont{J.}~\bibnamefont{Schrieffer}} \bibnamefont{and}
  \bibinfo{author}{\bibfnamefont{P.}~\bibnamefont{Wolff}},
  \bibinfo{journal}{Phys. Rev.} \textbf{\bibinfo{volume}{149}},
  \bibinfo{pages}{491} (\bibinfo{year}{1966}).

\bibitem[{\citenamefont{K{\"u}bler et~al.}(1983)\citenamefont{K{\"u}bler,
  William, and Sommers}}]{heusler-sx}
\bibinfo{author}{\bibfnamefont{J.}~\bibnamefont{K{\"u}bler}},
  \bibinfo{author}{\bibfnamefont{A.}~\bibnamefont{William}}, \bibnamefont{and}
  \bibinfo{author}{\bibfnamefont{C.}~\bibnamefont{Sommers}},
  \bibinfo{journal}{Phys. Rev. B} \textbf{\bibinfo{volume}{28}},
  \bibinfo{pages}{1745} (\bibinfo{year}{1983}).

\bibitem[{\citenamefont{Lee et~al.}(2002)\citenamefont{Lee, Rhee, and
  Harmon}}]{lee-chi}
\bibinfo{author}{\bibfnamefont{Y.}~\bibnamefont{Lee}},
  \bibinfo{author}{\bibfnamefont{J.}~\bibnamefont{Rhee}}, \bibnamefont{and}
  \bibinfo{author}{\bibfnamefont{B.}~\bibnamefont{Harmon}},
  \bibinfo{journal}{Phys. Rev. B} \textbf{\bibinfo{volume}{66}},
  \bibinfo{pages}{054424} (\bibinfo{year}{2002}).

\bibitem[{\citenamefont{Wilson et~al.}(2008)\citenamefont{Wilson, Gygi, and
  Galli}}]{wilson-chi1}
\bibinfo{author}{\bibfnamefont{H.}~\bibnamefont{Wilson}},
  \bibinfo{author}{\bibfnamefont{F.}~\bibnamefont{Gygi}}, \bibnamefont{and}
  \bibinfo{author}{\bibfnamefont{G.}~\bibnamefont{Galli}},
  \bibinfo{journal}{Physical Review B} \textbf{\bibinfo{volume}{78}},
  \bibinfo{pages}{113303} (\bibinfo{year}{2008}).

\bibitem[{\citenamefont{Wilson et~al.}(2009)\citenamefont{Wilson, Lu, Gygi, and
  Galli}}]{wilson-chi2}
\bibinfo{author}{\bibfnamefont{H.}~\bibnamefont{Wilson}},
  \bibinfo{author}{\bibfnamefont{D.}~\bibnamefont{Lu}},
  \bibinfo{author}{\bibfnamefont{F.}~\bibnamefont{Gygi}}, \bibnamefont{and}
  \bibinfo{author}{\bibfnamefont{G.}~\bibnamefont{Galli}},
  \bibinfo{journal}{Physical Review B} \textbf{\bibinfo{volume}{79}},
  \bibinfo{pages}{245106} (\bibinfo{year}{2009}).

\bibitem[{\citenamefont{Li et~al.}(2011)\citenamefont{Li, Hu, Yang, Johansson,
  and Vitos}}]{hu-x2}
\bibinfo{author}{\bibfnamefont{C.}~\bibnamefont{Li}},
  \bibinfo{author}{\bibfnamefont{Q.}~\bibnamefont{Hu}},
  \bibinfo{author}{\bibfnamefont{R.}~\bibnamefont{Yang}},
  \bibinfo{author}{\bibfnamefont{B.}~\bibnamefont{Johansson}},
  \bibnamefont{and} \bibinfo{author}{\bibfnamefont{L.}~\bibnamefont{Vitos}},
  \bibinfo{journal}{Appl. Phys. Lett.} \textbf{\bibinfo{volume}{98}},
  \bibinfo{pages}{261903} (\bibinfo{year}{2011}).

\bibitem[{\citenamefont{Chadov et~al.}(2009)\citenamefont{Chadov, Fecher,
  Felser, Min{\'a}r, Braun, and Ebert}}]{chadov-dftu}
\bibinfo{author}{\bibfnamefont{S.}~\bibnamefont{Chadov}},
  \bibinfo{author}{\bibfnamefont{G.}~\bibnamefont{Fecher}},
  \bibinfo{author}{\bibfnamefont{C.}~\bibnamefont{Felser}},
  \bibinfo{author}{\bibfnamefont{J.}~\bibnamefont{Min{\'a}r}},
  \bibinfo{author}{\bibfnamefont{J.}~\bibnamefont{Braun}}, \bibnamefont{and}
  \bibinfo{author}{\bibfnamefont{H.}~\bibnamefont{Ebert}}, \bibinfo{journal}{J.
  Phys. D: Appl. Phys.} \textbf{\bibinfo{volume}{42}}, \bibinfo{pages}{084002}
  (\bibinfo{year}{2009}).

\bibitem[{\citenamefont{Baroni et~al.}(2001)\citenamefont{Baroni, De~Gironcoli,
  Dal~Corso, and Giannozzi}}]{baroni01}
\bibinfo{author}{\bibfnamefont{S.}~\bibnamefont{Baroni}},
  \bibinfo{author}{\bibfnamefont{S.}~\bibnamefont{De~Gironcoli}},
  \bibinfo{author}{\bibfnamefont{A.}~\bibnamefont{Dal~Corso}},
  \bibnamefont{and}
  \bibinfo{author}{\bibfnamefont{P.}~\bibnamefont{Giannozzi}},
  \bibinfo{journal}{Rev. Mod. Phys.} \textbf{\bibinfo{volume}{73}},
  \bibinfo{pages}{515} (\bibinfo{year}{2001}).

\bibitem[{\citenamefont{Floris et~al.}(2011)\citenamefont{Floris, de~Gironcoli,
  Gross, and Cococcioni}}]{phononu}
\bibinfo{author}{\bibfnamefont{A.}~\bibnamefont{Floris}},
  \bibinfo{author}{\bibfnamefont{S.}~\bibnamefont{de~Gironcoli}},
  \bibinfo{author}{\bibfnamefont{E.}~\bibnamefont{Gross}}, \bibnamefont{and}
  \bibinfo{author}{\bibfnamefont{M.}~\bibnamefont{Cococcioni}},
  \bibinfo{journal}{Phys. Rev. B} \textbf{\bibinfo{volume}{84}},
  \bibinfo{pages}{161102} (\bibinfo{year}{2011}).

\end{thebibliography}

\end{document}